\documentclass[a4paper,fleqn,usenatbib]{mnras}

\usepackage{newtxtext,newtxmath}

\usepackage[T1]{fontenc}
\usepackage{ae,aecompl}

\usepackage{graphicx}	
\usepackage{amsmath}	
\usepackage{amssymb}	
\usepackage{url} 
\usepackage{comment}
\usepackage{subfig}

\def\Lya{Ly$\alpha$}
\def\Ha{H$\alpha$}
\def\Hb{H$\beta$}
\def\OII{[O\,{\sc ii}]}
\def\OIII{[O\,{\sc iii}]}
\def\OIIIuv{O\,{\sc iii}]}
\def\CIII{C{\sc iii}]}
\def\CIV{C{\sc iv}}
\def\HeII{He{\sc ii}}
\def\NV{N{\sc v}}

\def\xiion{$\xi_{\rm ion}$}
\def\MUV{$M_{\rm UV}$}
\def\DeltavLya{$\Delta v_{{\rm Ly}\alpha}$}
\def\ebv{$E(B-V)$}

\def\HI{H\,{\sc i}}

\def\ergsHz{erg$^{-1}$\,Hz}

\def\aj{AJ}
\def\apj{ApJ}
\def\apjs{ApJS}
\def\apjl{ApJL}
\def\aap{A\&A}	
\def\mnras{MNRAS}
\def\pasj{PASJ}

\title[UV spectroscopic properties of faint $z\sim 3$ LAEs]
{The Mean Ultraviolet Spectrum of a Representative Sample of Faint $z\sim 3$ Lyman Alpha Emitters}

\author[Nakajima et al.]{%
Kimihiko Nakajima, $^{1}$%
	\thanks{JSPS Overseas Research Fellow; E-mail: knakajim@eso.org}
Thomas Fletcher,$^{2}$
Richard S. Ellis,$^{1,2}$
Brant E. Robertson, $^{3}$ 
\newauthor 
and Ikuru Iwata $^{4}$
\\
$^1$European Southern Observatory, 
	Karl-Schwarzschild-Str. 2, D-85748, Garching bei M\"{u}nchen, Germany\\
$^2$Department of Physics and Astronomy, 
	University College London, Gower Street, 
	London WC1E 6BT, UK \\
$^3$Department of Astronomy and Astrophysics, 
	University of California, Santa Cruz, 
	1156 High Street, Santa Cruz, CA 95064, USA \\
$^4$Subaru Telescope, National Astronomical Observatory of Japan, 
	650 North A‘ohoku Place, Hilo, HI 96720, USA \\
}

\begin{document}

\label{firstpage}

\date{Accepted for publication in MNRAS on Mar 17, 2018}

\pagerange{\pageref{firstpage}--\pageref{lastpage}} \pubyear{2018}

\maketitle


\begin{abstract}

We discuss the rest-frame ultraviolet emission line spectra of a large 
($\sim 100$) sample of low luminosity redshift $z\sim 3.1$ Lyman alpha emitters (LAEs) drawn 
from a Subaru imaging survey in the SSA22 survey field. Our earlier 
work based on smaller samples indicated that such sources have high 
\OIII$/$\OII\ line ratios possibly arising from a hard ionising spectrum that 
may be typical of similar sources in the reionisation era. 
With optical spectra secured from VLT/VIMOS, we re-examine the 
nature of the ionising radiation in a larger sample using the strength of the high ionisation
diagnostic emission lines of  \CIII$\lambda1909$, \CIV$\lambda 1549$, \HeII$\lambda 1640$, and 
\OIIIuv$\lambda\lambda 1661,1666$ \AA\ in various stacked subsets. 
Our analysis confirms earlier suggestions of a correlation between the 
strength of Ly$\alpha$ and \CIII\ emission and we find similar trends with
broad band UV luminosity and rest-frame UV colour. Using various
diagnostic line ratios and our stellar photoionisation models, we determine both the
gas phase metallicity and hardness of the ionisation spectrum characterised
by \xiion\ - the number of Lyman continuum photons per UV luminosity.
We confirm our earlier suggestion that \xiion\ is significantly larger for LAEs than for
continuum-selected Lyman break galaxies, particularly for those LAEs
with the faintest UV luminosities. We briefly discuss the implications for
cosmic reionisation if the metal-poor intensely star-forming systems studied
here are representative examples of those at much higher redshift.

\end{abstract}

\begin{keywords}
Galaxies: evolution - galaxies: high-redshift
\end{keywords}


\section{Introduction} 
\label{sec:introduction}

A continuing debate in the high redshift community is whether early star-forming galaxies generate 
sufficient photons to govern the reionisation process. The key questions relate to both the nature of 
the ionising spectrum for low metallicity intensely-star forming systems thought to dominate the 
redshift interval $6<z<10$, and the extent to which the circumgalactic medium is transparent to 
Lyman continuum (LyC) radiation. Detailed diagnostics of these questions are currently challenging to 
observe for $z>6$ sources. 
The nature of the ionising spectrum, particularly the efficiency of the ionising photon production 
from the stellar population defined by \xiion, the number of LyC photons per UV luminosity 
(e.g., \citealt{robertson2013}), is ideally constrained by  measures of Balmer emission 
\citep{bouwens2016} interpreted through recombination physics. However, the Balmer emission lines
are currently beyond reach of ground-based telescopes at high redshifts. Furthermore, due to the 
rapid increase in intergalactic absorbers LyC leakage from galaxies at $z>6$ cannot be directly 
measured (e.g. \citealt{madau1995,II2008,inoue2014}).
For these reasons, attention has focused on identifying and studying possible analogues 
of low metallicity galaxies at 
low (e.g., \citealt{leitherer2016,izotov2016a,izotov2016b,%
schaerer2016,verhamme2017})
and intermediate redshifts 
(e.g., \citealt{iwata2009,nestor2013,vanzella2016a,vanzella2016b,amorin2017}).

\begin{table}
  \centering
  \caption{Summary of Optical Imaging Data.
    }
  \label{tbl:optical_imaging}
  \renewcommand{\arraystretch}{1.25}
  \begin{tabular}{@{}lccccc@{}}
    \hline
    Band &
    Field & 
    Observatory &
    PSF &
    $m_{\rm lim}$ &
    Reference \\
     &
     & 
     &
    {\scriptsize (1)} &
    {\scriptsize (2)} &
    {\scriptsize (3)} \\
    \hline
    $NB497$ &
    SSA22 &
    Subaru &
    $1.0$ &
    $26.2$ &
    (a), (b), (c) \\
    $B$ &
    SSA22 &
    Subaru &
    $1.0$ &
    $26.5$ &
    (a), (b), (c) \\
    $V$ &
    SSA22 &
    Subaru &
    $1.0$ &
    $26.6$ &
    (a), (b), (c) \\
    $R$ &
    SSA22 &
    Subaru &
    $1.1$ &
    $26.7$ &
    (a), (c) \\
    $i^{\prime}$ &
    SSA22 &
    Subaru &
    $1.0$ &
    $26.4$ &
    (a) \\
    $z^{\prime}$ &
    SSA22 &
    Subaru &
    $1.0$ &
    $25.7$ &
    (a) \\
    \hline
  \end{tabular}
  \\ 
  \vspace{-1mm}
  \begin{flushleft}
	(1) FWHM in arcsec.
	(2) $5\sigma$ limiting magnitude estimated by 
	$2^{\prime\prime}$ diameter aperture photometry.
	(3) (a) \citet{hayashino2004}; 
	(b) \citet{yamada2012}; 
	(c) \citet{matsuda2004}.
  \end{flushleft}
\end{table}

Low metallicity, low mass systems such as narrow-band selected Lyman alpha emitters (LAEs) at 
intermediate redshift may provide valuable examples of sources contributing to cosmic 
reionisation. In a recent paper \citep{nakajima2016} we have demonstrated for a modest sample 
of $15$ $z\sim 3$ LAEs that \OIII$\lambda\lambda 5007,4959$ emission is unusually intense 
compared to \OII$\lambda 3727$,  contrary to similar measurements of the more metal rich Lyman 
break population (LBGs).  We showed that this enhanced \OIII\ is 
unlikely to be the result of a lower metallicity alone and may arise, at least partially, from a harder 
ionising spectrum. Since intense \OIII\ emission appears to be a common feature of sources in 
the reionisation era, as inferred from \textit{Spitzer}/IRAC broad band photometry 
\citep{schenker2013,smit2014,smit2015,roberts-borsani2016}, we argued that intrinsically faint 
$z\sim 3$ LAEs may be valuable analogues of the dominant population of faint sources which may 
govern reionisation.
Their importance in understanding the early universe also arises from the fact that the
relative fraction of all star-forming galaxies showing strong Lyman alpha emission increases with 
redshift up to the reionisation era (e.g., \citealt{stark2011,curtis-lake2012,cassata2015}).

In this paper we attempt to further constrain the nature of the ionising radiation, 
i.e., the efficiency of the LyC photon production characterised by \xiion, 
for a much larger sample of $\sim 100$ $z\sim 3$ LAEs drawn from the same parent dataset as
in \citet{nakajima2016} via constraints on the strength of high ionisation UV metal lines, 
principally \CIII$\lambda 1909$ and \CIV$\lambda 1549$ which are redshifted into the optical at 
$z\simeq$3.1%
\footnote{
The \CIII$\lambda 1909$, or simply \CIII, refers to the sum of the 
$[$\CIII$\lambda 1907$ and \CIII$\lambda 1909$. 
The \CIV$\lambda 1549$, or \CIV, quotes the \CIV\ doublet of 
$\lambda=1548$\,\AA\ and $1550$\,\AA.
}.  This approach follows a similar analysis undertaken for sub-luminous lensed
Lyman break galaxies at $z\sim 2-3$ \citep{stark2014} interpreted with photoionisation codes 
(e.g., \citealt{NO2014,feltre2016,gutkin2016,nakajima2017}) which also supported the notion of a harder
ionising radiation field compared to local sources.

A plan of the paper follows. In \S2 we introduce the sample targeted
for follow-up spectroscopy which is drawn from a field previously surveyed
with narrow-band filters with the Subaru telescope. We discuss
the observations undertaken with both the Keck and ESO VLT telescopes
and the techniques used to reduce these data. In \S3 we discuss
the spectroscopic results in terms of the success rates of recovering
Ly$\alpha$ emission in the two observational campaigns and the 
velocity offset of Ly$\alpha$, $\Delta\,v_{Ly\alpha}$, relative to the systemic velocity inferred
from nebular lines. The relationship between $\Delta\,v_{Ly\alpha}$ and
the equivalent width (EW(Ly$\alpha$)) is then used to stack the higher quality ESO
spectra in various ways. Furthermore, we interpret these spectroscopic
stacks and provide constraints on the ratios of various UV metal lines which we interpret 
with photoionisation models thereby deriving measures of the gas phase metallicity
and hardness of the ionising spectrum. We discuss the implications of our
results and present a summary in \S4.

We assume a $\Lambda$CDM cosmology using the Planck results
with $\Omega_M$=0.308, $\Omega_{\Lambda}$=0.692 and $H_0$=67.8
km s$^{-1}$ Mpc$^{-1}$. 
We adopt a solar chemical composition following \citet{asplund2009}.

\begin{figure}
	\centerline{
    		\includegraphics[bb=0 0 568 396, width=0.99\columnwidth]{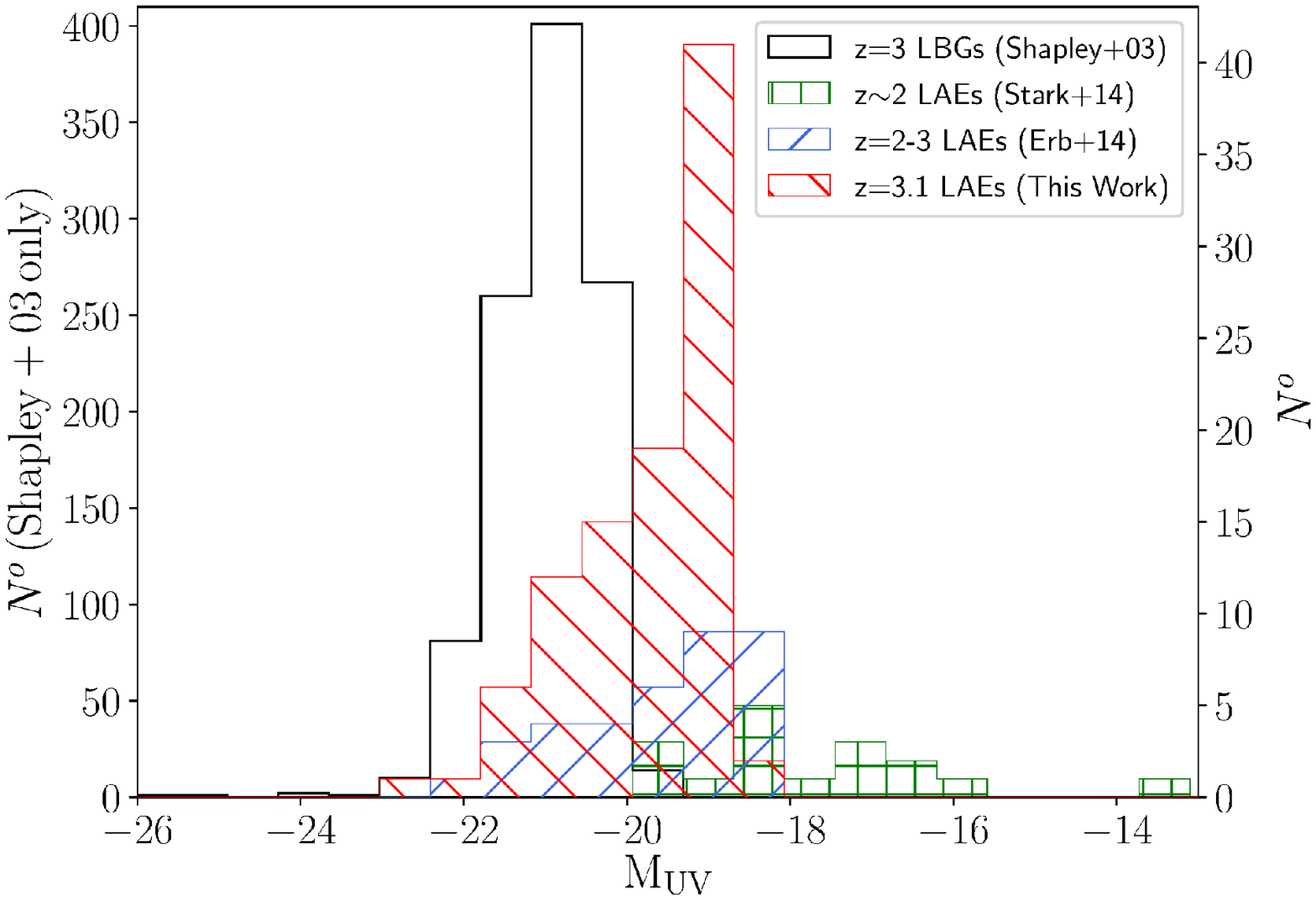}
  	}
  	\caption{
		UV absolute magnitude distributions for recently-published rest-frame UV spectroscopic 
		studies at $z=2-4$. Our VIMOS sample is plotted with the red hatched histogram. Other 
		datasets include Lyman break galaxies: \citet{shapley2003} in black (numbers in the left 
		ordinate) and \citet{erb2014} in blue, and gravitationally-lensed galaxies from 
		\citet{stark2014} in green.
 	}
	\label{fig:histo_MUV}
\end{figure}

\begin{table*}
  \centering
  \caption{Summary of the VIMOS observation for the $z=3.1$ LAEs.
    }
  \label{tbl:observations_summary}
  \renewcommand{\arraystretch}{1.25}
  \begin{tabular}{@{}lcclcccc@{}}
    \hline
    Instrument &
    Field & 
    Date of obs. &
    Exposure times &
    Wavelength coverage &
    $R$ &
    \multicolumn{2}{|c|}{$N^o$ of LAEs [LBGs]$^{(\dag)}$} 
    \\
     &
     &
     &
    (hrs) &
    (\AA) &
     &
    Observed &
    Identified \\ 
    \hline
    VIMOS &
    SSA22 &
    2016 Aug \& Oct &
    $11.0$ (quads. \#1--3) &
    4850 -- 9450 &
    580 &
    $72$ $[10]$ &
    $59$ $[1]$$^{(\ddag)}$ \\
     &
     &
     &
    $7.0$ (quad. \#4)&
     &
     &
    $12$ $[3]$ &
    $6$ $[1]$ \\
    \hline
  \end{tabular}
  \\ 
  \vspace{-1mm}
  \begin{flushleft}
  	\small
  	($\dag$) Numbers of LBGs are given in the square brackets.
	($\ddag$) Three additional LBGs that lack \Lya\ emission are identified with UV absorption lines.
  \end{flushleft}
\end{table*}

\section{Spectroscopic Data} 
\label{sec:data}

\subsection{Sample} 
\label{ssec:data_sample}

Our target sample of $z\sim 3.1$ Lyman alpha emitters (LAEs) is drawn from earlier panoramic 
narrow-band imaging surveys undertaken with the Subaru Telescope. In our previous paper (\citealt{nakajima2016}, hereafter Paper I), we secured near-infrared spectra sampling \OII, \Hb\ and 
\OIII\ emission for $15$ sources in the SSA22 field \citep{hayashino2004,yamada2012,micheva2017}
using Keck/MOSFIRE. $13$ of these $15$ objects are LAEs with EW(\Lya) $>20$\,\AA\ and the 
remaining two are Lyman break galaxies (LBGs) also at $z\simeq 3.1$. The present enlarged 
sample is drawn from $281$ LAEs in the same field (SSA22-Sb1; \citealt{yamada2012}) 
including the 13 MOSFIRE-identified LAEs
selected via their photometric excess in a narrowband
filter centred at $497$\,nm.  In order to obtain rest-frame UV spectra to further our studies of the 
nature of this population, we conducted deep optical spectroscopy of a subset of these 
photometrically-identified  LAEs using the VLT/VIMOS spectrograph. Optical spectroscopy was also 
obtained using the Keck/LRIS spectrograph in a second SXDS field for which MOSFIRE data is 
available (Fletcher et al 2018, in prep). That data is only used here to enlarge the sample for which 
Ly$\alpha$ velocity offsets can be determined (\S\ref{ssec:results_lya}).

\begin{figure*}
    \begin{center}
        \subfloat{
            \includegraphics[width=0.4\textwidth]{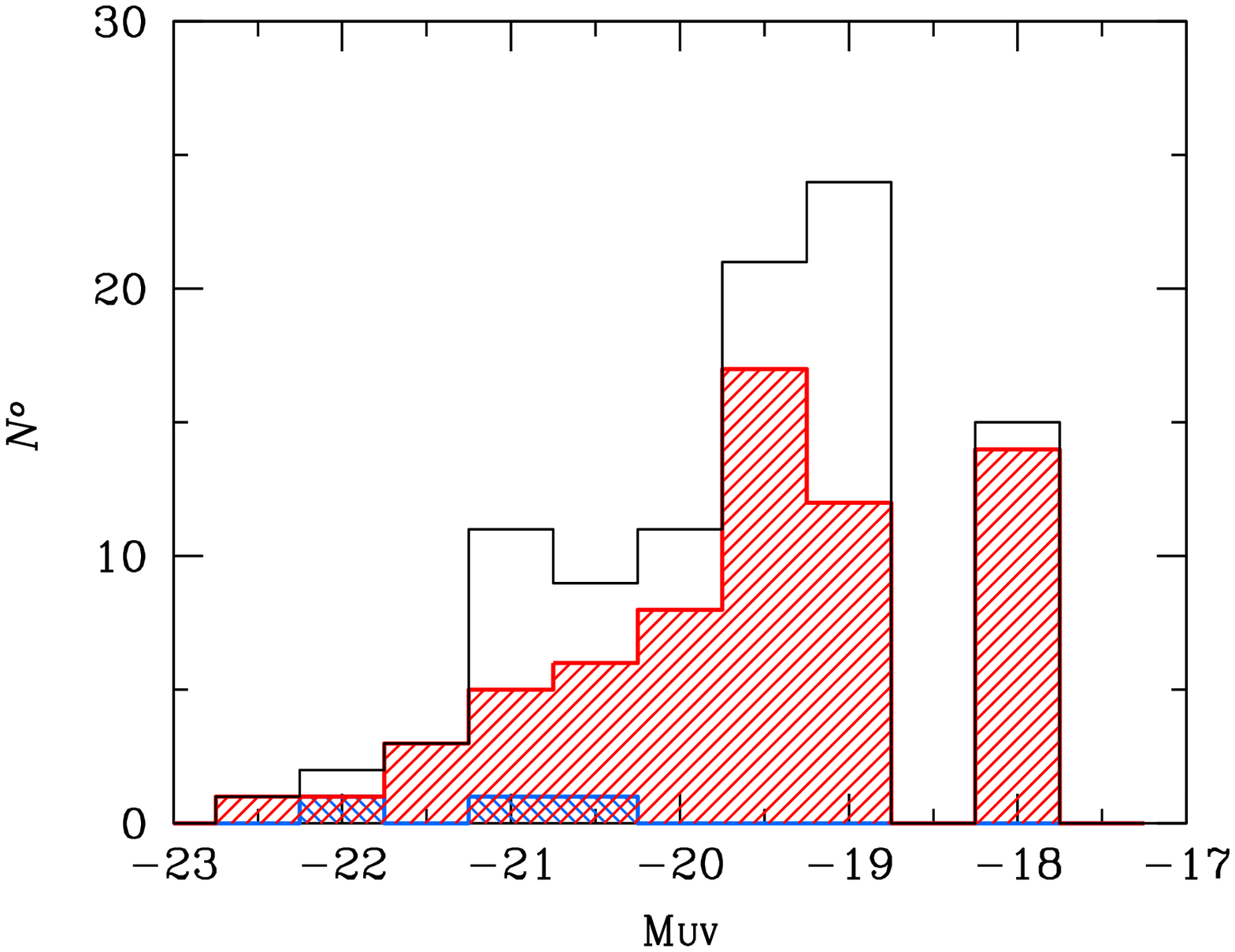}
        }
        \subfloat{
            \includegraphics[width=0.4\textwidth]{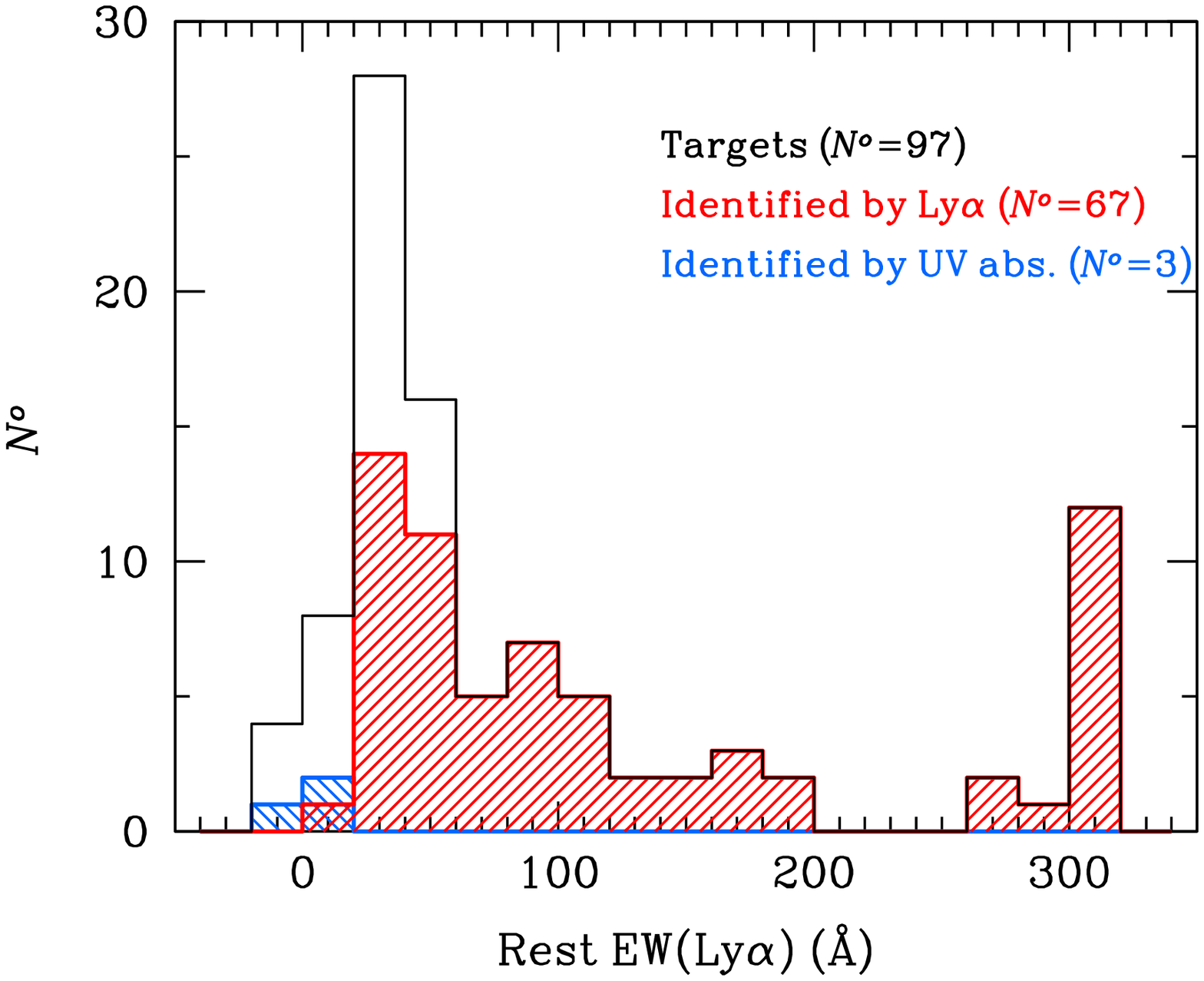}
        }
        \caption{%
        		Distributions of \MUV\ (left) and EW(\Lya) (right)
		for the VIMOS sample. 
		Red hatched histograms show the numbers of galaxies 
		whose \Lya\ is spectroscopically confirmed, and the blue histogram presents LBGs 
		identified via UV absorption line(s).
		LAEs whose UV continuum was not detected in the Subaru images
		are assigned to \MUV\ $> -18.5$ and EW(\Lya) $>300$\,\AA.
        }
        \label{fig:lya_success}
    \end{center} 
\end{figure*}

In the VIMOS sample we secured UV spectra for $7$ of the $13$ LAEs and one of the two LBGs in the 
SSA22 field for which MOSFIRE spectra are available from Paper I. Additionally, we sampled a further 
$77$ LAEs and $12$ LBGs. In total, our new enlarged sample of rest-frame optical spectroscopy 
comprises $84$ LAEs and $13$ LBGs, representing a considerable advance over the sample discussed 
in Paper I. Finally, we have likewise recently enlarged the sample for which MOSFIRE near-infrared 
spectra is available (Nakajima et al 2018, in prep). This data provides new rest-frame optical emission 
lines for $6$ LAEs and $2$ LBGs and is also only used here to improve our understanding of the 
Ly$\alpha$ velocity offsets (\S\ref{ssec:results_lya}). Further details of the observations are given in 
the following subsection.

An important advantage of selecting the SSA22 field is the plentiful deep multi-wavelength photometric 
data. The optical broadband data is useful in constraining the nature of the stellar populations, such as 
the absolute UV magnitude \MUV\ and the UV continuum slope parameter $\beta$. 
We derive the absolute UV magnitude using $R$-band photometry which traces the rest frame spectral 
energy distribution (SED) around $1500$--$1600$\,\AA, and the UV continuum slope $\beta$ is 
measured assuming a power-law fit to the $R$, $i$, and $z$-band photometry according to
the convention $f_{\lambda}\propto \lambda^{\beta}$. Such $\beta$ measurements are thus only 
possible for those LAEs detected in at least two of the Subaru $R$, $i$, and $z$-band images. 
A resulting distribution of the UV slope is $\beta \simeq -1.6 \pm 0.8$, which corresponds to \ebv\ 
$\simeq0.09\pm0.07$ with the SMC extinction curve and the BPASS SEDs \citep{reddy2018}.
This is comparable to the results of other photometric studies of LAEs at similar redshifts, where 
$\beta$ values are calculated as in \citet{meurer1997} (e.g., \citealt{nilsson2009}) and \ebv\ via 
SED fitting analysis (e.g., \citealt{ono2010,nilsson2011,guaita2011,kusakabe2018}).
Finally, although the VIMOS spectra are crucial in confirming the Ly$\alpha$ emission, 
they cannot provide accurate measures of the equivalent width (EW) of Ly$\alpha$ due to the 
low signal-to-noise of the UV continuum and aperture effects inevitable with slit spectroscopy
(e.g., \citealt{momose2016,wisotzki2016}). It is preferable, therefore, to
measure EWs of \Lya\ from the BV $-$ NB497 colours, where BV stands for a 
combination of the B and V broad-band images as ($2$B$+$V)$/3$ \citep{hayashino2004}. 
This can be done by using the \Lya\ observed wavelength in the EW calculation and taking into account
the narrrow-band and broad-band filter transmission curves to accurately 
translate the BV $-$ NB497 colour into the EW.
The limiting depths of the relevant broad-band photometry in the SSA22 field are 
given in Table \ref{tbl:optical_imaging}.

It is useful to compare the properties of our new spectroscopic sample with those for which similar
line emission studies have been undertaken. Fig.\ \ref{fig:histo_MUV} compares the UV luminosity 
distribution of our $97$ targets with those for other $z\simeq 2-3$ star-forming galaxies 
\citep{shapley2003,erb2014,stark2014}. Our LAE sample probes a \MUV\ range from $-22.4$ down to 
$-18.7$ with a median value of $-19.5$. Notably, $15$ of our LAEs are fainter than the broad-band
detection limit (\MUV\ $\gtrsim -18.7$) and the $800$ or so UV-selected LBGs from \citet{shapley2003} 
with $R\sim24.4$ -- $24.9$, corresponding to \MUV\ of $-21.1$ to $-20.6$. 
Only the $16$ gravitationally-lensed galaxies at $z=1.6-3$ studied by \citet{stark2014} overlaps
our luminosity range. That sample extends from \MUV\ $=-19.9$ down to $-13.7$.

\subsection{VIMOS Observations} 
\label{ssec:data_observation}

The VIMOS spectroscopic observations were undertaken in the SSA22 field on UT3 of
the VLT in Service Mode (ID: 098.A-0010(A), PI; Ellis) in August  -- October 2016. A single VIMOS
pointing was adopted, comprising a slit mask across four quadrants of the instrument with a field of view of 
$4\times 7^{\prime}\times 8^{\prime}$. To mitigate atmospheric dispersion, the slitlet position angle is
restricted to PA $=90^{\circ}$ and the field centre was chosen to maximise the number of targets for
which \OIII$\lambda 5007$ had been detected already with MOSFIRE (Paper I).
Seven LAEs and one LBG from Paper I fell in one quadrant (\# 3) and the remaining sample
totalling $97$ sources included $77$ new LAEs at $z=3.1$ and $12$ LBGs at $z=3-4$
as well as the $8$ MOSFIRE objects. The new sample of $77$ LAEs was drawn uniformly from the 
parent catalogue and has a median \MUV\ of $-19.5$.

We observed in medium-resolution mode with the GG475 order sorting filter which provided a 
wavelength coverage from $\sim 4800$\,\AA\ to $\sim 1\,\mu$m at a mean 
spectral resolution of $R\sim 580$ with a slit width of $1^{\prime\prime}$.
Our observation comprised $13$ identical observing blocks 
(OBs), each with an on-source integration  of $3\times 20$\,min 
using a three point offset pattern along the slit. Relatively bright stars were 
included on each quadrant adopted to monitor sky conditions and mask alignment.  
Two of the 13 OBs were discarded as the seeing inferred was $\gtrsim 1^{\prime\prime}$.  
For quadrant \#4, the first 5 OBs, one of which was under a poor seeing condition,
were lost due to either an improper alignment 
or mis-insertion of the milled mask. The other three quadrants in these 5 OBs 
were unaffected, since mask alignment was performed excluding the alignment 
stars in quadrant \#4. We thus discarded 4 additional OBs' data for quadrant \#4. 

In summary, therefore, we obtained in total $11$ hrs of on-source integration data 
for objects in the quadrants \#1, 2, and 3, and $7$ hrs in the quadrant \#4.
The typical seeing for the acceptable OBs was $0\farcs 75$.

The VIMOS data was reduced using recipes in the standard ESO VIMOS pipeline (v.3.1.7) 
operated through esoreflex (v.2.8.5). For each OB, the process included 
bad pixel cleaning,  bias subtraction, flat-field correction, frame combination, 
sky fringing correction, wavelength calibration, and flux calibration.
We performed wavelength calibration and flat field using relevant
frames taken after each OB.
Using the slit position of each science target output by the VIMOS pipeline,
we extracted the individual 2D spectrum. Sky subtraction was performed with 
a low-order polynomial fit along the slit for each wavelength.
The 2D spectra were corrected for Galactic extinction 
based on the \ebv\ map of \citet{schlegel1998}.
Finally, we combined the individual sky-subtracted 2D spectra 
($N=11$ for objects in the quadrants \#1, 2, and 3, and $N=7$ in the quadrant \#4) 
with a sigma-clipped average to produce a final 2D spectrum 
for each of the science targets.
1D spectra were produced for those targets for which at least one emission line 
could be identified (normally Ly$\alpha$) adding up $9$ pixels along the spatial direction 
centred on the detected emission line. This width was chosen to maximise the signal-to-noise 
ratio and corresponds approximately to twice the seeing size. 

\begin{figure*}
    \begin{center}
        \subfloat{
            \includegraphics[width=0.49\textwidth]{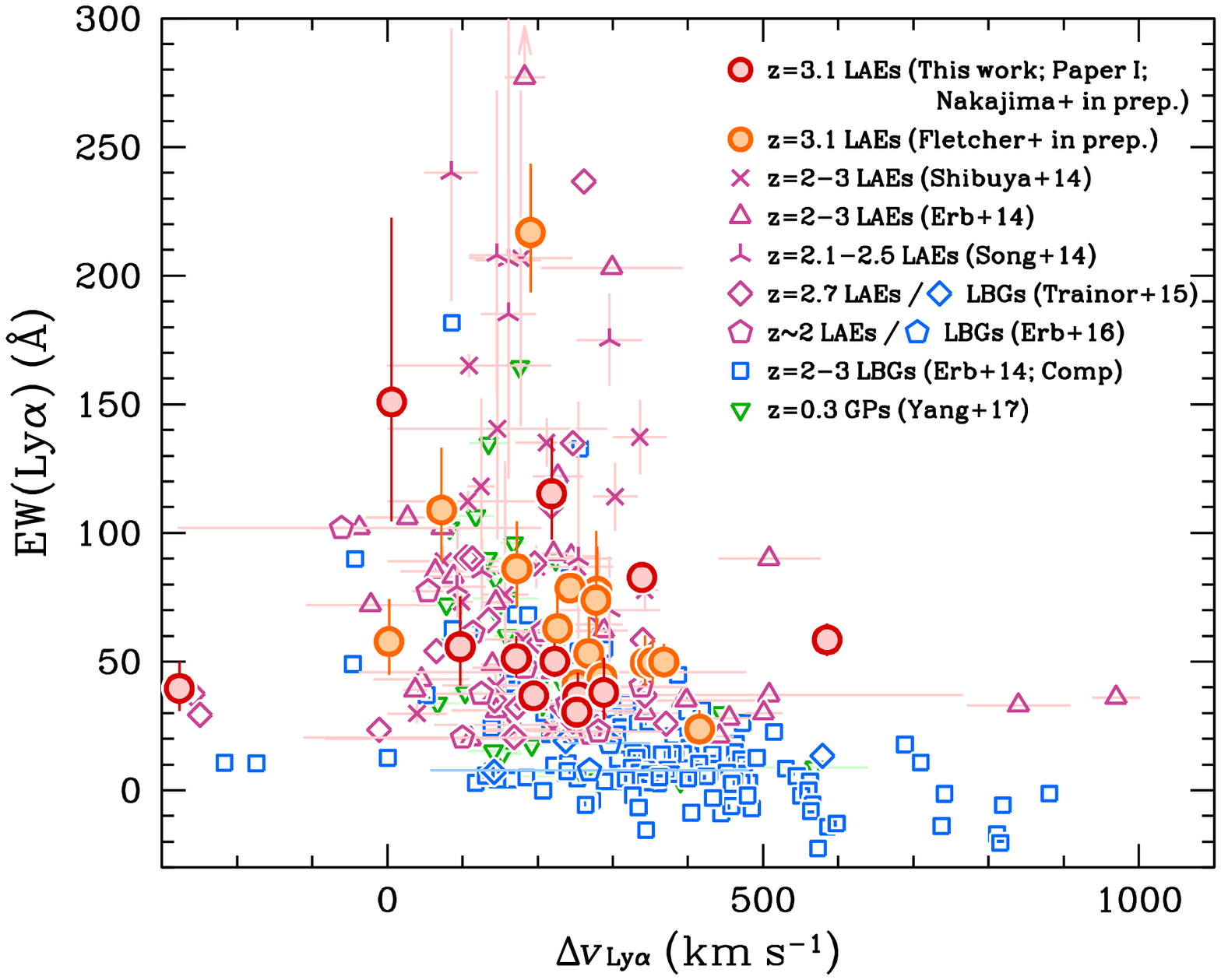}
        }
        \subfloat{
            \includegraphics[width=0.49\textwidth]{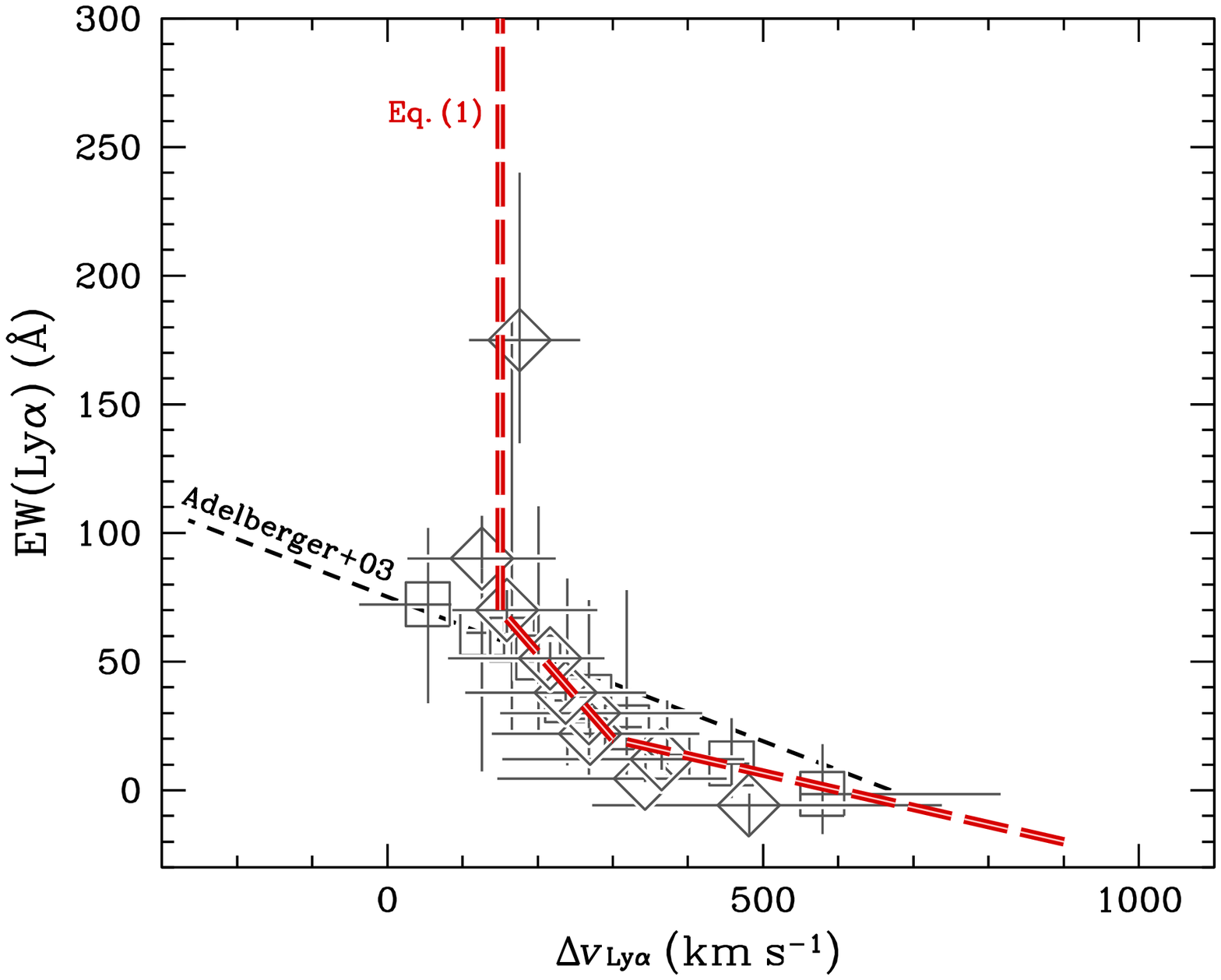}
        }
        \caption{%
        		(Left:) 
		Relationship between EW(\Lya) and the velocity offset \DeltavLya\ 
		for the VIMOS and other samples. 
		Red filled circles represent the newly observed LAEs 
		with VIMOS whose systemic redshift is 
		determined using MOSFIRE detections of [O III] (Paper I). 
		Orange circles show those additional LAEs 
		observed with LRIS and recent MOSFIRE data (see text for details).
		Magenta, blue and green symbols represent equivalent measures for high redshift
		LAEs, LBGs with weak \Lya\ (EW $<20$\,\AA),and 
		low-$z$ green pea galaxies, respectively (see legend).
		(Right:)
		Open grey squares present the average EW(\Lya)
		for a \DeltavLya\ bin, and open grey diamonds
		show the average \DeltavLya\ for a EW(\Lya) bin, 
		using all the individual data points shown in the left panel.
		The red long-dashed lines is the derived relationship 
		based on Eq. (\ref{eq:vLya_ewLya}), while the black dashed 
		line shows that from \citet{adelberger2003}.
        }
        \label{fig:vLya_ewLya}
    \end{center} 
\end{figure*}


\subsection{Spectroscopic Confirmations}
\label{ssec:data_detections}

Our initial task is to spectroscopically confirm the validity of the photometric identifications
of Ly$\alpha$ emission at the expected redshift of $z\simeq3.1$. Earlier
work drawn from the same Subaru narrow-band imaging survey \citep{hayashino2004}
has found a typical success rate of $\simeq 66$\,\%\ \citep{matsuda2005} (see \S\ref{ssec:results_lya}). 
Initially therefore, two authors independently examined the VIMOS spectra for detectable Ly$\alpha$ 
emission through visual inspection of both the 1-D and 2-D spectra separately.
Such an approach was considered adequate given that the wavelength of \Lya\ 
emission is accurately predicted by the narrowband filters.

Applying a 3$\sigma$ signal to noise limit for the line detections, this
investigation resulted in a confirmed list of $65$ $z=3.1$ LAEs corresponding to 
a relative success rate is $77$\,\%\ ($=65/84$).
Figure \ref{fig:lya_success} shows the success rate as a function of \MUV\ and EW(\Lya).
Notably, we recover \Lya\ emission spectroscopically for $100$\,\%\ of those LAEs with 
EW(\Lya) $>60$\,\AA. 
The success rate is higher than reported by \citet{matsuda2005} ($66$\,\%).
We consider this reasonable because (i) our VIMOS observations are deeper than 
earlier studies, and (ii) our sample probes fainter sources.
Among the $16$ objects studied with MOSFIRE, both in Paper I
and with our most recent campaign (Nakajima et al. 2018, in prep.), $12$ show \Lya\ emission in 
the VIMOS spectrum.

From our VIMOS observations we also find $2$ out of $13$ LBGs show prominent 
\Lya\ emission with rest-frame EW of $>20$\,\AA. Hereafter we will refer to these $2$ 
LBGs and the $65$ genuine LAEs observed as the ``VIMOS-LAEs''. 
We are confident that such sources cannot be explained as lower-$z$ emitters, 
such as \OII$\lambda 3727$ emitters at $z\simeq 0.3$ and \Hb-or-\OIII\ emitters at $z\lesssim 0.03$ 
because of the lack of stronger emission lines of \OIII$\lambda 5007$ and \Ha\ at the 
corresponding wavelengths.

At this stage, prior to stacking the VIMOS spectra (\S\ref{ssec:results_stacked_spectra}), we also 
inspected the individual spectra searching for other emission lines. 
Among the $67$ VIMOS-LAEs, only $15$ sources reveal additional UV metal lines including: 
\CIII$\lambda 1909$,
\OIIIuv$\lambda 1665$,
\HeII$\lambda 1640$, and/or 
\CIV$\lambda 1549$\,\AA.
Most of these sources ($10/15$) only present \CIII\ emission 
with a tentative detection of the other UV line(s). These sources are generally 
the brighter subset in the spectroscopic sample, including one galaxy initially selected 
as a LBG and one LAE known as a TypeII AGN (Paper I).  We additionally find $3$ LBGs 
whose \Lya\ emission is not visible but for which UV absorption lines are detected. 

We can now summarise the fruits of our spectroscopic campaign for a total of $97$ targets
(see Table \ref{tbl:observations_summary}). The catalogue summarising which sources have
individual metal line detections is provided in Table \ref{tbl:properties_individuals}.

\begin{figure*}
	\centerline{
    		\includegraphics[width=0.85\textwidth]{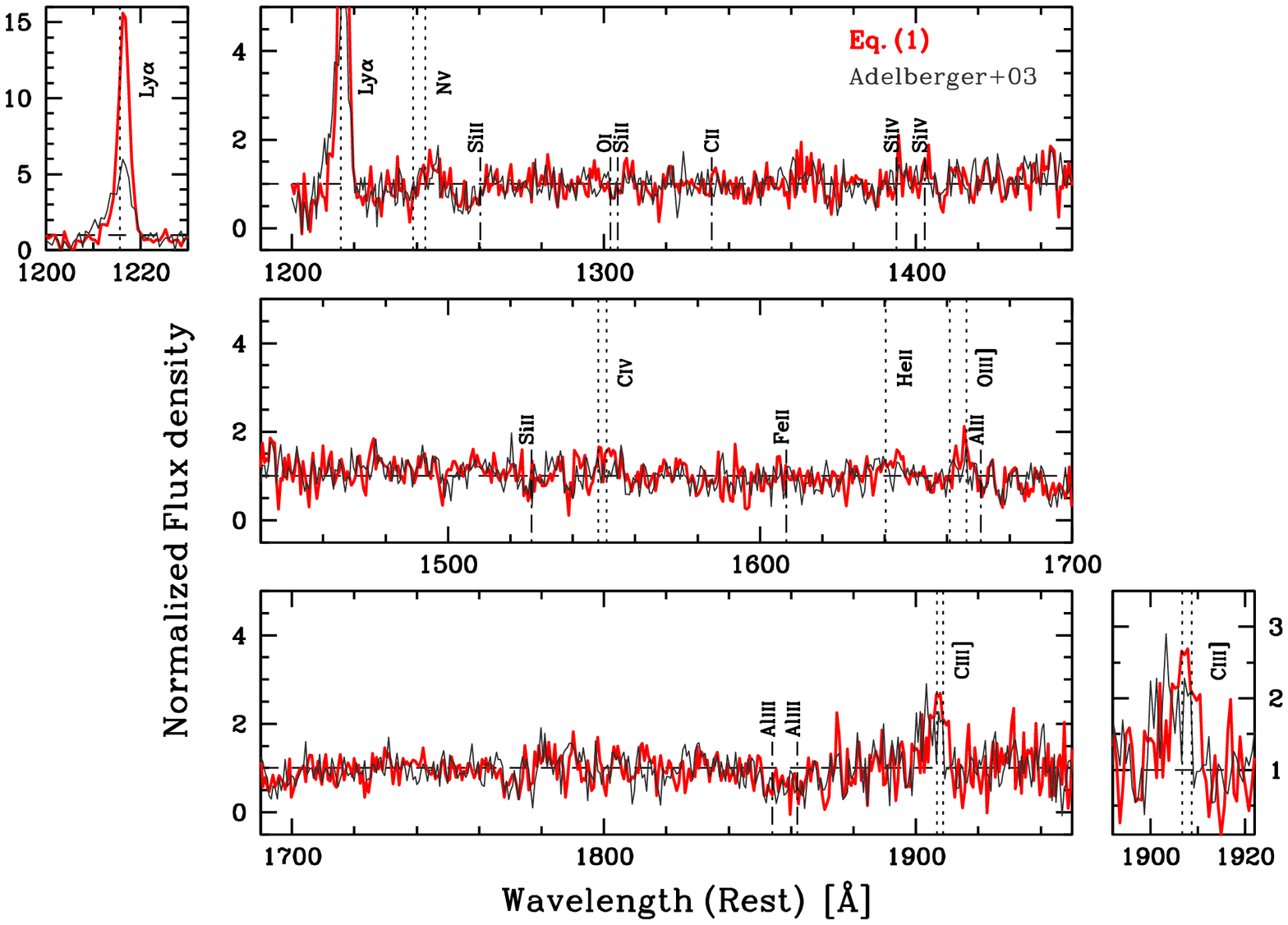}
  	}
  	\caption{
		The normalised composite rest-frame spectrum for the sample of $70$ 
		spectroscopically-confirmed VIMOS LAEs.
		The thick red spectrum adopts the relationship between \DeltavLya\ and 
		EW(\Lya) that we derive with the latest compilation of LAEs and LBGs
		(i.e., Eq. (\ref{eq:vLya_ewLya})), while the thin black is based on the 
		\citet{adelberger2003} relation.	Additional panels focus on the \Lya\ and \CIII\ emission
		regions and highlight the difference between the two methods.
		Wavelengths of additional emission lines and absorption lines
		are marked with a dotted and dot-dashed vertical line,
		respectively. 
 	}
	\label{fig:compare_spec_entire}
\end{figure*}

\section{Results} 
\label{sec:results}

\subsection{\Lya\ emission and its velocity offset} 
\label{ssec:results_lya}

The overlap of $12$ confirmed LAEs with our earlier MOSFIRE data (Paper I plus Nakajima et al. 
2018, in prep.) 
and our new Ly$\alpha$ detections from both VIMOS and LRIS (Fletcher et al 2018, in prep)
enables us to revisit the question of the velocity offset of \Lya, \DeltavLya, with respect
to other nebular lines. This offset is crucial to understand prior to stacking spectra in order to achieve 
reliable detection of weaker UV metal lines. Traditionally this quantity has been investigated as 
a function of the EW of \Lya\ \citep{adelberger2003} and we examine this relationship anew 
combining both literature data 
\citep{shibuya2014,erb2014,song2014,trainor2015,erb2016,yang2017}
and our own SSA22 measures in Figure \ref{fig:vLya_ewLya}. As discussed in 
\S\ref{ssec:data_sample} and above, we bring into play both the LRIS and recent MOSFIRE data 
only for the purpose of increasing the sample size.

Examination of this data suggests a new relationship given by: 
\begin{eqnarray}
	\begin{array}{l}
		\Delta v_{{\rm Ly}\alpha}\,({\rm km\,s}^{-1}) \\
		\\
		\\
	\end{array}
	\begin{array}{ll}
		  \simeq \ 150 	& ({\rm EW}>70\,\mbox{\AA}) \\
		  \simeq \ 360 - 3\,{\rm EW} 	& ({\rm EW}=20-70\,\mbox{\AA}) \\
		  \simeq \ 600 - 15\,{\rm EW} 	& ({\rm EW}<20\,\mbox{\AA}). 
	\end{array}
	\label{eq:vLya_ewLya}	
\end{eqnarray}
Such an anti-correlation between EW(\Lya) and $\Delta v_{{\rm Ly}\alpha}$
has been previously reported in datasets with a wide dynamic range in EW(\Lya)
(e.g., \citealt{hashimoto2013,shibuya2014,erb2014,trainor2015}).
As discussed by earlier kinematic studies of LAEs, the column density 
of \HI\ gas must play a key role in shaping the anti-correlation, with a larger 
offset from systems of a higher column density.
By combining all the individual data points from the literature, 
we confirm that the correlation becomes steeper toward a smaller 
$\Delta v_{{\rm Ly}\alpha}$, i.e., that the spread in EW distribution 
increases with a smaller \Lya\ velocity offset. 
This trend is as theoretically expected (e.g., \citealt{ZW2014}), as 
the secondary effects of anisotropy of the system and viewing angle
become less significant in a higher column density.
We cannot however fully examine the relationship for LAEs with 
EW(\Lya) $\gtrsim 100$\,\AA\ due to the modest sample size, and 
thus present an average velocity offset of $\Delta v_{{\rm Ly}\alpha}\simeq 150$\,km\,s$^{-1}$
for such strong LAEs. The 
kinematics properties of our LAEs will be discussed in more detail elsewhere.
The relation of Eq. (\ref{eq:vLya_ewLya}) appears to be more appropriate than that presented by 
\citet{adelberger2003}, especially for galaxies showing strong \Lya\ emission. 
We investigate the difference between both relationships in our stacking analysis below.

\begin{figure*}
  \centering
  \subfloat{%
    \includegraphics[width=0.85\textwidth]{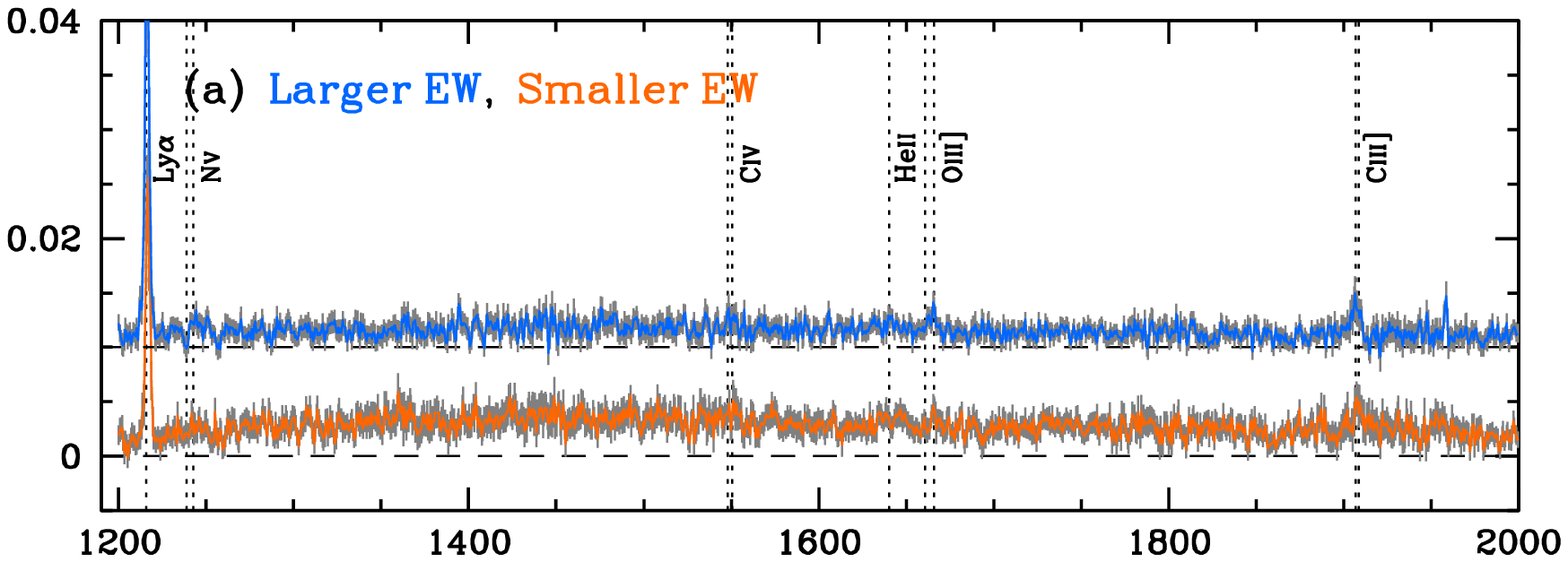}%
  }\      \subfloat{%
    \includegraphics[width=0.85\textwidth]{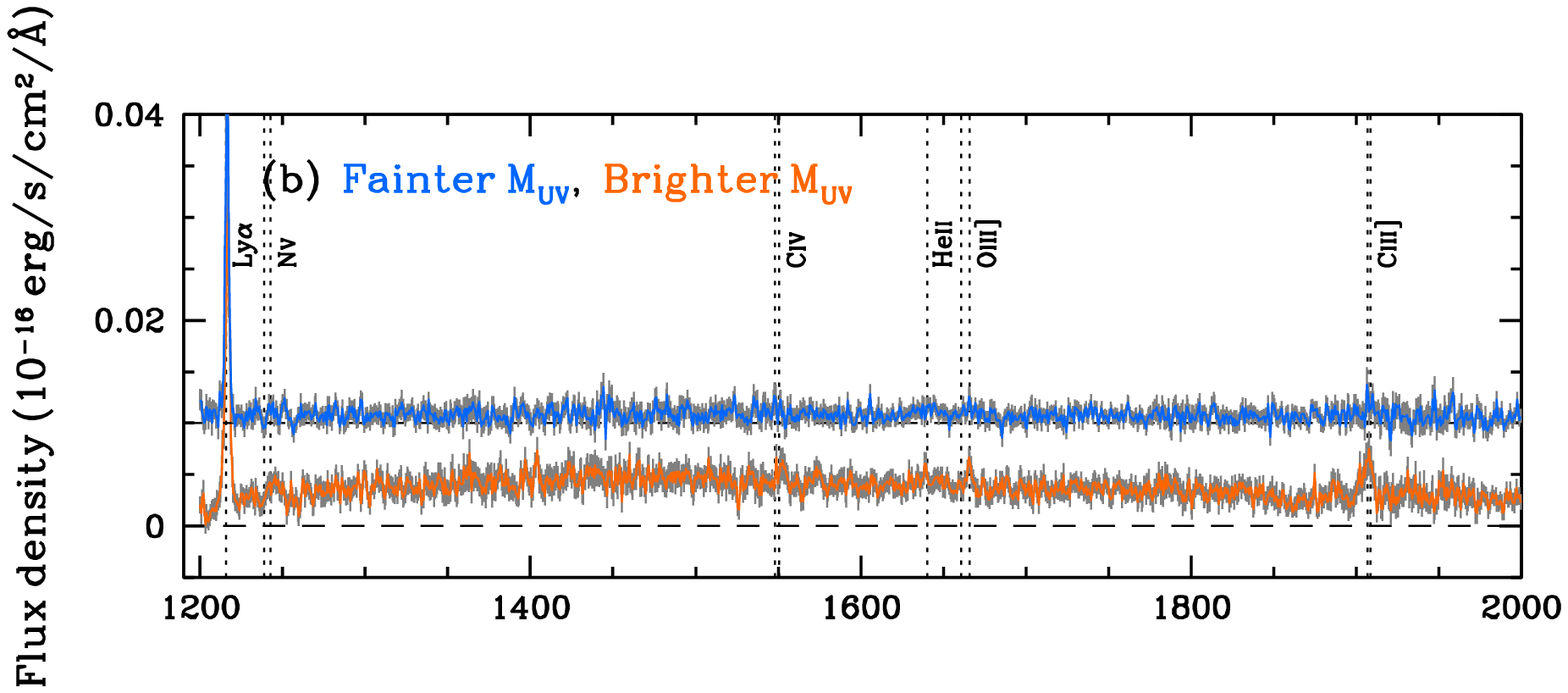}%
  }\      \subfloat{%
    \includegraphics[width=0.85\textwidth]{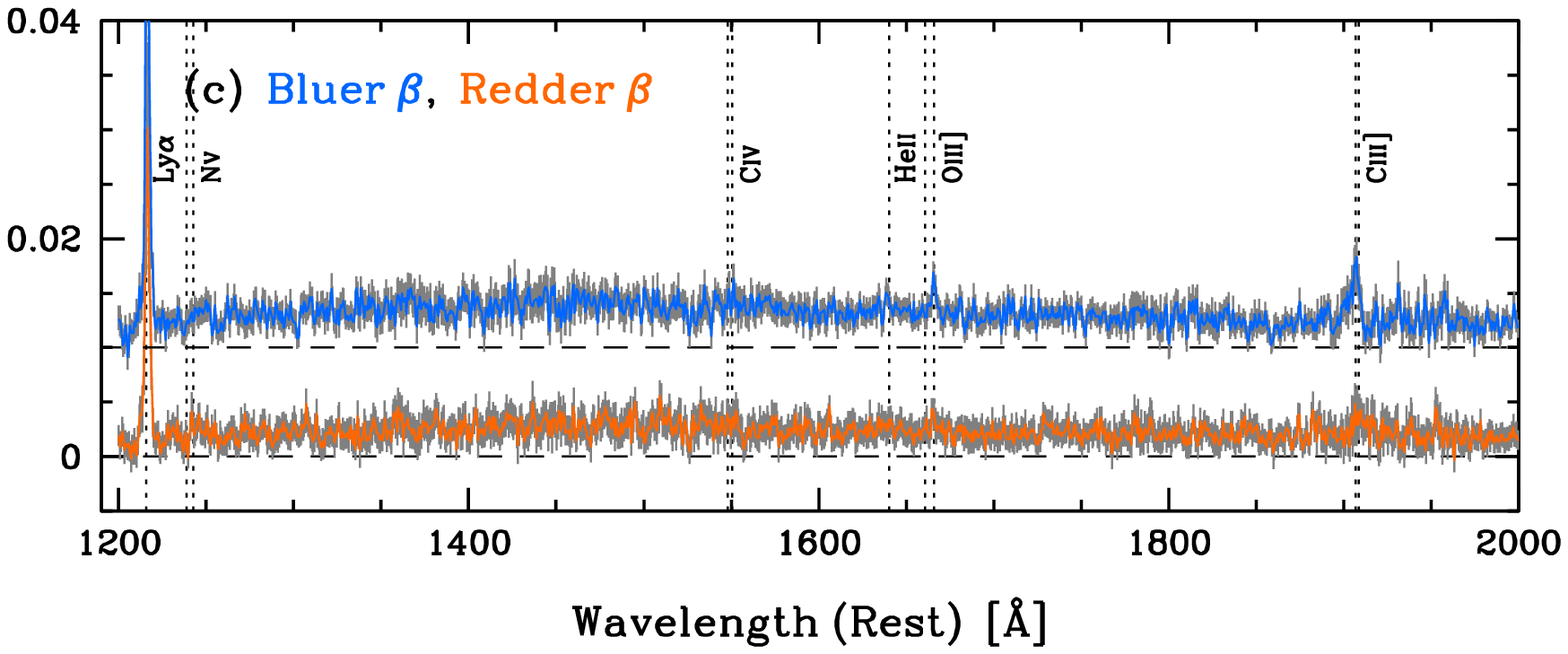}%
  }
  \caption{
  		Composite rest-frame spectra of the VIMOS LAEs separated according to the strength of
		(a) EW(\Lya), (b) \MUV, and (c) UV slope $\beta$.
		In each case the smaller/fainter and larger/brighter subsamples are displayed in blue/orange respectively
		 in each panel. For convenience the blue spectrum is offset vertically.
		The grey shaded region around each spectrum refers to the standard deviation 
		of the flux density at each wavelength estimated by bootstrap resampling (see text for details).
		The wavelengths of key diagnostic emission lines are marked with a
		dotted line.
  }
  \label{fig:spec_subsamples}
\end{figure*}

\subsection{Stacked Spectra} 
\label{ssec:results_stacked_spectra}

Despite our significant integration time, we only directly detect other UV emission lines in a small 
subset of our data (\S \ref{ssec:data_detections}, Table \ref{tbl:properties_individuals}).
To determine the average strengths of \CIII\ and other diagnostic UV metal lines we adopted the 
following stacking procedure.

Firstly, we removed the single known AGN (\S\ref{ssec:data_detections}) from the sample 
leaving $66$ VIMOS-LAEs. Additionally, since their rest-frame optical lines are identified, we 
included those $4$ targets confirmed with MOSFIRE for which \Lya\ is not visible in the VIMOS 
spectrum. The sample available for stacking analyses thus consists of $70$ sources. 

Using the individual flux-calibrated spectra, we shifted each to the rest-frame as described below, 
rebinned each to a common dispersion of $0.65$\,\AA\ per pixel, and scaled to a common median 
in the wavelength range of $\sim 1250$ -- $1500$\,\AA. Finally, the spectra were averaged. 
To exclude positive and negative sky subtraction residuals and cosmic-ray residuals, 
we rejected an equal number of the highest and lowest outliers at each wavelength corresponding 
in total to $\sim 5$\,\%\ of the data. We confirmed that median-stacked composite spectra
generated in the same way were almost indistinguishable from the average-stacked ones.

To define a systemic redshift for each galaxy, we adopted a similar approach used by 
\citet{shapley2003}. We used the rest-frame optical redshifts for the MOSFIRE identified objects in 
SSA22 (Paper I; Nakajima et al. in prep.). For the others, we estimated a systemic redshift from 
the \Lya\ using the updated formula given by Eq. (\ref{eq:vLya_ewLya}).

The normalised composite spectrum for the entire sample is shown in red in Figure 
\ref{fig:compare_spec_entire}; that adopted using \citet{adelberger2003}'s velocity offset
prescription is shown in black. The most striking difference is found around the \Lya\ emission. 
The red spectrum presents a prominent, asymmetric line. On the other hand, the black shows 
a less peaked signal with a broad tail to the blue side of the emission. This originates from the simple 
linear relation of \citet{adelberger2003}, \DeltavLya\ $\simeq 670-8.9 {\rm EW}$, whereby 
a fraction of the VIMOS LAEs with EW $\gtrsim 75$\,\AA\ would be predicted to have a blueshifted 
\Lya\ emission with respect to the systemic redshift.
We can also evaluate the validity of the composite method by measuring the EWs of \Lya.
We determine the EW(\Lya) from the black and red spectrum as $31\pm 8$\,\AA\ and $67\pm 17$\,\AA,
respectively. Given the typical EW(\Lya) of $73$\,\AA\ estimated from the narrow-band photometry
in the VIMOS-LAEs sample, we consider the red spectrum using Eq. (\ref{eq:vLya_ewLya}) to be 
the more accurate prescription.
Additionally, the sought-after diagnostic UV lines of \CIII, \CIV, and \OIIIuv\ are detected with 
greater significance in the red composite spectrum, whilst the noise level around these emission 
lines is greater in the black spectrum. Figure \ref{fig:compare_spec_entire} thus demonstrates 
that the prescription of Eq. (\ref{eq:vLya_ewLya}) is preferred for the \DeltavLya\ estimation 
from EW(\Lya), especially for LAEs with a large EW. 

To evaluate sample variance and statistical noise, we adopt a bootstrap technique, similar to that 
used in \citet{shapley2003}. We generated $1000$ fake composite spectra constructed from the 
sample of spectra used in creating the real composite spectra. Each fake spectrum was constructed 
in the same way, with the same number of spectra as the actual composite, but with the list of input 
spectra formulated by selecting spectra at random, with replacement, from the full list of VIMOS LAE 
spectra. With these $1000$ fake spectra, we derived the standard deviation at each dispersion pixel. 
The standard deviations are taken into account in the following analyses based on line fluxes and 
EWs.

\begin{table*}
  \centering
  \caption{Spectroscopic properties of subsamples 
    }
  \label{tbl:measurements_subsamples}
  \renewcommand{\arraystretch}{1.25}
  \begin{tabular}{lcccccc}
    \hline
     &
    Smaller EW(\Lya) &
    Larger EW(\Lya) &
    Brighter \MUV\ &
    Fainter \MUV\ & 
    Redder $\beta$ &
    Bluer $\beta$ \\
    \hline
    $N_{\rm gal}$ &
    $35$ &
    $35$ &
    $35$ &
    $35$ &
    $25$ &
    $26$ \\
    $\langle {\rm EW}({\rm Ly}\alpha)\rangle$ &
    $38.$ &
    $119.$ &
    $40.$ &
    $113.$ &
    $42.$ &
    $63.$ \\
    $\langle M_{\rm UV}\rangle$ &
    $-20.00$ &
    $-19.20$ &
    $-20.32$ &
    $-18.97$ &
    $-19.72$ &
    $-19.69$ \\
    $\langle \beta\rangle$ &
    $-1.5$ &
    $-2.3$ &
    $-1.5$ &
    $-3.3$ &
    $-1.2$ &
    $-2.0$ \\
    \CIV\ EW &
    $1.41 \pm 0.53$ &
    $2.91 \pm 0.77$ &
    $1.75 \pm 0.45$ &
    $3.88 \pm 1.18$ &
    $<1.57$ &
    $1.61 \pm 0.47$ \\
    \CIV$/$\CIII\ &
    $0.36 \pm 0.16$ &
    $0.29 \pm 0.09$ &
    $0.30 \pm 0.08$ &
    $0.49 \pm 0.19$ &
    $<0.39$ &
    $0.22 \pm 0.07$ \\
    \CIV$/$\CIII$_{\rm corr}$$^{(\dag)}$ &
    $0.46 \pm 0.20$ &
    $0.32 \pm 0.09$ &
    $0.38 \pm 0.11$ &
    $0.49 \pm 0.19$ &
    $<0.54$ &
    $0.25 \pm 0.08$ \\
    \HeII\ EW &
    $1.15 \pm 0.35$ &
    $1.83 \pm 0.45$ &
    $1.26 \pm 0.30$ &
    $2.03 \pm 0.53$ &
    $1.12 \pm 0.34$ &
    $1.52 \pm 0.38$ \\
    \HeII\ EW$_{\rm corr}$$^{(\ddag)}$ &
    $0.15 \pm 0.05$ &
    $0.83 \pm 0.26$ &
    $0.26 \pm 0.08$ &
    $1.03 \pm 0.34$ &
    $0.11 \pm 0.04$ &
    $0.51 \pm 0.17$ \\
    \HeII$/$\CIII\ &
    $0.26 \pm 0.10$ &
    $0.19 \pm 0.05$ &
    $0.20 \pm 0.05$ &
    $0.26 \pm 0.09$ &
    $0.22 \pm 0.09$ &
    $0.19 \pm 0.05$ \\
    \HeII$/$\CIII$_{\rm corr}$$^{(\dag)}$$^{(\ddag)}$ &
    $0.039 \pm 0.020$ &
    $0.089 \pm 0.038$ &
    $0.050 \pm 0.020$ &
    $0.130 \pm 0.062$ &
    $0.029 \pm 0.016$ &
    $0.073 \pm 0.031$ \\
    \OIIIuv\ EW &
    $1.89 \pm 0.66$ &
    $4.67 \pm 0.87$ &
    $1.99 \pm 0.30$ &
    $3.60 \pm 1.32$ &
    $1.68 \pm 0.41$ &
    $2.95 \pm 0.36$ \\
    \OIIIuv$/$\CIII\ &
    $0.42 \pm 0.17$ &
    $0.45 \pm 0.10$ &
    $0.31 \pm 0.06$ &
    $0.39 \pm 0.17$ &
    $0.34 \pm 0.12$ &
    $0.37 \pm 0.07$ \\
    \OIIIuv$/$\CIII$_{\rm corr}$$^{(\dag)}$ &
    $0.49 \pm 0.20$ &
    $0.47 \pm 0.11$ &
    $0.36 \pm 0.07$ &
    $0.39 \pm 0.17$ &
    $0.41 \pm 0.15$ &
    $0.41 \pm 0.07$ \\
    \CIII\ EW &
    $5.41 \pm 1.20$ &
    $15.14 \pm 1.99$ &
    $9.17 \pm 0.98$ &
    $11.05 \pm 2.62$ &
    $5.89 \pm 1.57$ &
    $10.99 \pm 1.41$ \\
    \hline
  \end{tabular}
  \\ 
  \vspace{-1mm}
  \begin{flushleft}
  	\small
  	($\dag$) Corrected for reddening. 
	($\ddag$) Corrected for the \HeII\ stellar emission 
	(\S \ref{sssec:results_stacked_spectra_corrections}).
  \end{flushleft}
\end{table*}

\begin{figure}
	\centerline{
    		\includegraphics[width=0.95\columnwidth]{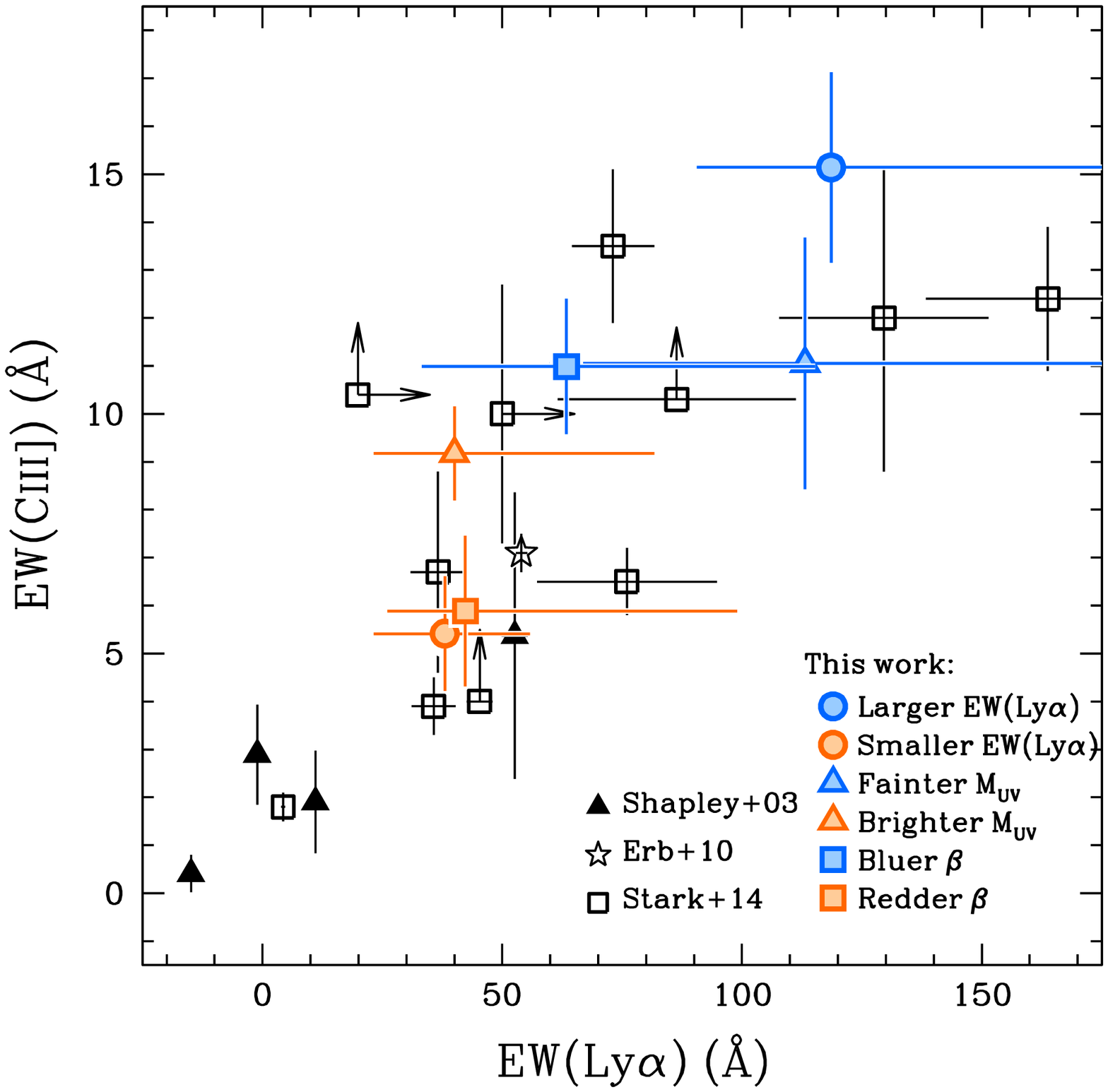}
  	}
  	\caption{
		Relationship between EW(\Lya) and EW(\CIII)
		for the VIMOS and other samples. 
		The blue and orange filled symbols show 
		the composites of the six VIMOS subsamples
		as indicated in the legend. 
		The black symbols refer to measures for $z=2-3$ galaxies 
		compiled from the literature 
		as described in the legend.
	}
	\label{fig:ewc3_ewlya}
\end{figure}

\subsubsection{Dependencies}
\label{sssec:results_stacked_spectra_dependencies}

To examine spectroscopic trends in the LAE sample, we now split the dataset into various subsamples 
based on the properties of the individual spectra. Since most of the LAEs are faint, we simply divide 
the sample into two halves to maximise the S$/$N of the resulting composite spectra. We chose the 
key properties of the EW(\Lya), absolute UV magnitude \MUV, and UV continuum slope $\beta$
to investigate the various trends. 
The median and the threshold values are EW(\Lya) $= 73$\,\AA, \MUV\ $=-19.5$, and $\beta = -1.50$.
Note that the subsamples selected according
to the UV slope are somewhat smaller since some LAEs lack the necessary
multiple bands of rest-frame photometry. The composite spectra of the subsamples are shown in 
Figure \ref{fig:spec_subsamples}, and the various properties derived from the composite spectra are 
listed in Table \ref{tbl:measurements_subsamples}.  The errors on these entries take into account the 
variance measure on the stacked spectra as detailed above.

We begin by discussing Figure \ref{fig:ewc3_ewlya} which presents the correlation 
between EW(\Lya) and EW(\CIII) for the subsamples. A clear trend is obvious
and follows that previously reported \citep{stark2014,rigby2015,lefevre2017}. 
\CIII\ emission is stronger in LAEs with larger EW(\Lya), fainter \MUV, and bluer UV slopes. 
Similar trends are also observed in \CIV, \OIIIuv, 
and \HeII\ emission (corrected; \S\ref{sssec:results_stacked_spectra_corrections}).
As clarified by earlier studies with photoionisation models \citep{JR2016,nakajima2017}, 
simple stellar populations find it hard to reproduce \CIII\ with EW $\gtrsim 10$\,\AA\
and populations including the contribution of massive binary systems have been
proposed (\S \ref{sec:discussion}).

\subsubsection{Corrections to the nebular spectra}
\label{sssec:results_stacked_spectra_corrections}

To ensure the intrinsic nebular spectra of the VIMOS LAEs can be directly compared 
with photoionisation models (\S\ \ref{ssec:UV_diagnostics}), 
we must apply several corrections to the observed UV spectra.

Firstly, we need to correct for the contribution of stellar emission.
The \HeII\ emission line is a composite of both stellar and 
nebular emission (e.g. \citealt{brinchmann2008,erb2010,steidel2016}).
Since our VIMOS data do not have the spectral resolution necessary
to resolve the two components, we use a stellar synthesis code 
to predict the stellar contribution and hence subtract it. 
We use publicly available BPASS (v2.0; \citealt{stanway2015}) SEDs
including binary evolution to predict the stellar \HeII\ emission strengths. 
Considering a current star-formation age of $10$ to several $100$\,Myr
as plausible for LAEs (e.g. \citealt{gawiser2006,gawiser2007,ono2010,kusakabe2018}), 
the stellar \HeII\ emission is predicted to have EW $\simeq 0.8$ -- $1.4$\,\AA%
\footnote{%
The stellar \HeII\ emission strength is estimated to be 
EW $\simeq 0.8$ -- $1.2$\,\AA\ if we adopt the v2.1 BPASS 
models \citep{eldridge2017}. Our results are thus not affected 
by the choice between v2.0 and v2.1. In this paper, we use the v2.0 
BPASS models which are also adopted in the photoionisation models of 
\citet{nakajima2017} (\S\ref{ssec:UV_diagnostics}).
}.
As a fiducial value, we adopt a value of $1.0\pm 0.2$\,\AA, as used by \citet{nakajima2017}.
We consider this appropriate for a sub-solar metallicity of $Z\simeq 0.2-0.5\,Z_{\odot}$
(e.g., \citealt{nakajima2013,trainor2016}).
If a lower metallicity were assumed, the stellar \HeII\ emission could be 
slightly larger, $1.2\pm 0.2$\,\AA. 

A further correction is required for dust reddening. In Paper I
we assumed such a correction is negligible for LAE. Here, in an improvement,
we utilise the UV slope $\beta$ obtained from the Subaru photometry
in conjunction with our stellar population models.
Using the SMC extinction curve and the BPASS SEDs \citep{reddy2018},
we correct for the reddening of the line ratios for each of the  
six subsamples using the median value of $\beta$. 
The corrections are small and do not affect our results significantly. 
As an example, for the subsample with weaker EW(\Lya) $\beta \simeq 1.5$
and the corrected \CIV$/$\CIII\ ratio (which spans the widest wavelength
range we use) is increased by a factor of $\sim 1.3$ which is within the 
$1\sigma$ uncertainties. Although it is debatable whether nebular emission 
and the stellar continuum suffer the same amount of dust attenuation (e.g. \citealt{reddy2015}),
such differences will not be a concern for measures based on EWs.

\subsection{UV diagnostic diagrams} 
\label{ssec:UV_diagnostics}

\begin{figure*}
	\centerline{
    		\includegraphics[width=0.95\textwidth]{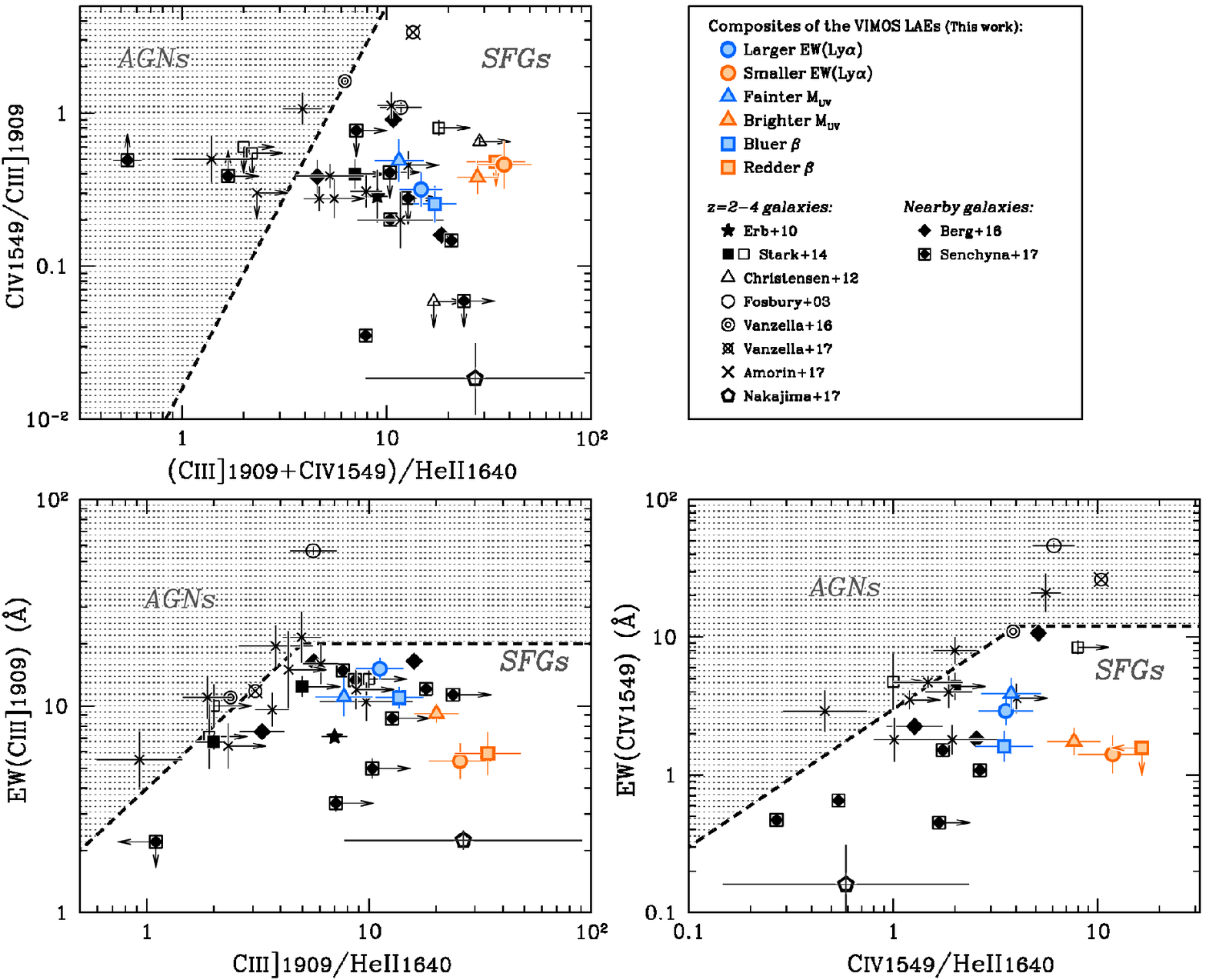}
  	}
  	\caption{
		Diagnostic diagrams involving various UV emission line ratios: 
		\CIV$/$\CIII\ vs. (\CIII$+$\CIV)$/$\HeII\ (top left), 
		EW(\CIII) vs. \CIII$/$\HeII\ (bottom left), and 
		EW(\CIV) vs. \CIV$/$\HeII\ (bottom right).
		Blue and orange symbols present the six 
		VIMOS subsamples as shown in the legend. 
		The black symbols are $z=0-4$ star-forming galaxies 
		compiled from the literature by \citet{nakajima2017},
		as shown in the legend.
		In these diagrams, star-forming galaxies can be separated 
		from AGNs, which are dominated in the grey hatched areas
		(see \citealt{nakajima2017}).
	}
	\label{fig:UV_diagnostics}
\end{figure*}

Following the corrections adopted in \S\ref{sssec:results_stacked_spectra_corrections}, we now 
proceed to interpret the UV spectra of our various VIMOS subsamples using three UV diagnostic 
emission line ratios in Figure \ref{fig:UV_diagnostics}. 
Since these are high ionisation lines, UV line diagnostics such as \CIII, \CIV\ and \HeII\
 were initially applied to spectra of radio galaxies and AGNs 
(e.g., \citealt{villar-martin1997,allen1998,groves2004,nagao2006,dors2014}). There is now broader
interest in studying these lines to gauge the ionisation field in star-forming galaxies, both 
theoretically (e.g., \citealt{feltre2016,gutkin2016,nakajima2017}) and 
observationally (e.g., \citealt{stark2014,amorin2017,lefevre2017}).
Our goal is to derive the physical properties of the gaseous phase in the interstellar medium (ISM)
in the context of the hardness of the ionising spectrum, as characterised by \xiion\ -- the number
of Lyman continuum photons per UV luminosity.  

In the following we follow the procedures introduced by \citet{nakajima2017}. In that work UV 
diagnostic diagrams are interpreted with photoionisation models which include incident radiation fields 
of both stellar populations and AGN and cover a wide range of metallicities and ionisation parameters. 
The authors demonstrate how such diagrams can be used not only to separate star-forming galaxies 
from AGN but also to constrain \xiion\ for star-forming galaxies. Fortunately, as expected, 
Figure \ref{fig:UV_diagnostics} demonstrates that all of our various LAE subsamples are consistent 
with star-forming regions. Thus, in the following, we only use the stellar photoionisation models to 
discuss their properties. 
We briefly describe the details of the stellar photoionisation models below.

Our stellar photoionisation models are based on \textsc{Cloudy} (version 13.03; 
\citealt{ferland1998,ferland2013}) and assume constant density homogeneous 
inter-stellar gas with a plane-parallel geometry. Dust physics and elemental depletion factors 
follow the analyses of \citet{dopita2006} and \citet{nagao2011}. All elements except nitrogen, carbon 
and helium are taken to be primary nucleosynthesis elements. 
For carbon and nitrogen, we use the formalism of \citet{dopita2006} and \citet{lopez-sanchez2012}, 
respectively, to account for their secondary products. For helium, we adopt the form 
given by \citet{dopita2006}. The photoionisation models are calculated by varying the
metallicity and ionisation parameter, fixing the gas density $n_{\rm H}=10^2$\,cm$^{-3}$.
For the radiation field, we adopt the population synthesis code 
BPASS (v2.0; \citealt{stanway2015}) for binary star populations and \textsc{PopStar} \citep{molla2009} 
for single stellar populations. Stellar metallicities are matched to their gas-phase equivalents. 
We assume a constant star-formation history and vary the duration of current star-formation
from $1$ to $1000$\,Myr. An older age corresponds to a lower \xiion\ parameter. 
Emission line ratios are insensitive to the duration of star-formation, but  
the EWs of UV lines are affected as the continuum level strongly depends on the older stellar population. 
The advantage of using the models of \citet{nakajima2017} is that they predict EWs 
of the UV lines as a function of metallicity, ionisation parameter, and \xiion.

We evaluate the average ionisation parameter and metallicity by fitting stellar
photoionisation models to the most sensitive line ratios of \CIV$/$\CIII\ and 
(\CIII$+$\CIV)$/$\HeII. The estimation procedure is similar to that typically adopted for 
the optical \OIII$/$\OII--(\OII$+$\OIII)$/$\Hb\ diagram (e.g. \citealt{KD2002,KK2004,NO2014})
but here applied to more highly ionised nebular regions. 
The (\CIII$+$\CIV)$/$\HeII\ is sensitive to metallicity and increases toward a higher 
metallicity. This is because of (i) an increase of carbon element and (ii) a softening of the 
radiation field (i.e., a weakening of the \HeII\ emission) with increasing metallicity.
The \CIV$/$\CIII\ is mainly governed by the ionisation parameter and increases with it. 
Moreover, the (\CIII$+$\CIV)$/$\HeII\ (\CIV$/$\CIII) ratio also has a weak dependency on 
the ionisation parameter (metallicity). We therefore use both the ratios of 
(\CIII$+$\CIV)$/$\HeII\ and \CIV$/$\CIII\ to estimate the metallicity and ionisation parameter 
simultaneously.
The top left panel of Figure \ref{fig:UV_diagnostics} shows that LAEs with larger EW(\Lya), 
fainter \MUV, and bluer UV slopes (hereafter called the ``stronger LAEs sample'') tend to 
present a smaller (\CIII$+$\CIV)$/$\HeII\ ratio. This indicates that the stronger LAEs sample 
has a lower gas-phase metallicity -- a result that would not be affected if we adopted a stronger 
correction for stellar emission of EW(\HeII)$_{\rm stellar}$ $=1.2\pm 0.2$\,\AA\ 
(\S \ref{sssec:results_stacked_spectra_corrections}). Based on the binary stellar population 
photoionisation models, the stronger LAEs sample has an average metallicity of  
$Z\simeq 0.05$ -- $0.2\,Z_{\odot}$, while the weaker LAEs sample has a somewhat 
higher metallicity of $Z\simeq 0.1$ -- $0.5\,Z_{\odot}$. 
The ionisation parameters $U$ -- defined as the ratio of the ionising photon flux to the gas density --
are constrained at the same time, differ with $\log U \simeq -1.5$ to $-2.2$ for the stronger 
LAEs sample compared to $\log U \simeq -2.0$ to $-2.5$ for the weaker LAEs sample. 
We derive the best-fit and $1\sigma$ uncertainty 
of metallicity and ionisation parameter for each of the three sub-samples in the stronger and 
weaker LAEs samples. The ranges of the properties above are given to cover the properties
of the three sub-samples including the $1\sigma$ uncertainties.
Although the subsample of LAEs with the redder UV slope does not exhibit a significant \CIV\ 
detection, its relatively large EW(\CIII) ensures that its ionisation parameter could not be less 
than $\log U \simeq -2.5$. These quantities are not significantly altered if we adopt the single 
stellar population models.

One uncertainty in the analysis is the use of \CIV\ as a diagnostic tool.
Since \CIV\ line photons can be trapped in a highly ionised nebular region by resonant scattering, 
they would be preferentially absorbed and weakened by internal dust. Although dust physics
is included in the modelling, there remain uncertainties in accurately modelling how this affects \CIV\ emission. 
For this work, we find that the LAEs are on average fairly free from dust with \ebv\ $=0.09\pm 0.07$ 
(\S\ref{ssec:data_sample}) and so assume \CIV\ measures will not be severely affected.

\begin{table*}
  \centering
  \caption{Summary of the inferred properties of the VIMOS LAEs with the UV lines.}
  \label{tbl:properties_UV}
  \renewcommand{\arraystretch}{1.25}
  \begin{tabular}{lccccc}
    \hline
    Sample &
    Input radiation &
    $Z$ ($Z_{\odot}$)&
    $\log U$ &
    Age (Myr) &
    $\log\,$\xiion$/$\ergsHz\ \\
    \hline
    Stronger LAEs &
     \multicolumn{5}{l}{\it (Subsamples of larger EW(\Lya), fainter \MUV, bluer UV slope)} \\
     &
     binary stellar pop. &
     $0.05$ $\cdots$ $0.2$ &
     $-2.2$ $\cdots$ $-1.5$ &
     $\lesssim 20$ &
     $25.68\pm 0.13$ \\
    \\
    Weaker LAEs &
     \multicolumn{5}{l}{\it (Subsamples of smaller EW(\Lya), brighter \MUV, redder UV slope)} \\
     &
     binary stellar pop. &
     $0.1$ $\cdots$ $0.5$ &
     $-2.5$ $\cdots$ $-2.$ &
     $10$ $\cdots$ $500$ &
     $25.54 \pm 0.09$ \\
     \hline
  \end{tabular}
\end{table*}

It is important to note that these models also reproduce our 
\OIIIuv\ emission line strengths. Since the stellar photoionisation models of \citet{nakajima2017}
by default assume an empirical relationship between C$/$O and 
O$/$H ratios \citep{dopita2006} to give a carbon abundance at 
each metallicity, LAEs are on average expected to obey the 
same relationship, as typically inferred for continuum-selected
galaxies at similar redshifts (e.g. \citealt{steidel2016}).
Using the method of \citet{PA2017}, the C$/$O ratios are 
estimated from the (\CIII$+$\CIV)$/$\OIIIuv\ ratios to be 
$\log$ C$/$O $\simeq -0.7\pm 0.1$.

\subsection{The Hardness of the Ionising Spectrum}
\label{ssec:xiion}

Finally, with the knowledge of the metallicity and ionisation parameter $U$, 
we turn to estimating the hardness of the ionising spectrum, \xiion. 
We consider both the EW(\CIII) and EW(\CIV) diagnostic diagrams presented
in the two lower panels of Figure \ref{fig:UV_diagnostics}.

The strength of the various UV lines is sensitive not only to
the ISM condition but also to the ratio of the number of ionising photons 
to the (non-ionising) UV continuum, i.e., \xiion. The UV nebular continuum must 
also be considered for an accurate analysis. Using the derived ISM properties 
(\S\ref{ssec:UV_diagnostics}) and assuming a constant star-formation rate, we vary the duration of
current star-formation from $1$ to $1000$\,Myr in order to match the observed 
EWs of \CIII\ and \CIV. The inferred age can then be directly translated into 
a constraint on \xiion. It is important to note that in the following derivation of
\xiion\  we will follow the convention of assuming a zero escape fraction of ionising photons
(see discussion in \S\ref{sec:discussion}). 

The derivation of \xiion\ also depends on whether or not the stellar population 
includes binary stars. As clarified by \citet{nakajima2017}, EWs of \CIII\ and \CIV\ generated
by a single stellar population produce EW(\CIII) $\lesssim 12$\,\AA\ and EW(\CIV) $\lesssim 9$\,\AA, 
whereas binary stellar populations yield EWs as large as $\sim 20$\,\AA\
and $\sim 12$\,\AA, respectively (see also \citealt{JR2016}).
Since the stronger LAEs sample shows EW(\CIII)s of $\simeq 13$\,\AA, a single stellar population is likely insufficient. 
We therefore adopt the binary stellar population models for
the stronger LAEs sample. 

After some consideration, we decided to focus on using the 
EW(\CIII) figure (lower left in Figure \ref{fig:UV_diagnostics} to estimate \xiion.
There remains a concern that the \CIV\ measurements may be affected by 
underlying stellar absorption. For the stronger LAEs sample, the large EW(\CIII) 
is consistent with a young star-formation age and a high \xiion. 
The inferred age spans a few Myr in the lowest metallicity case
($Z\simeq 0.05\,Z_{\odot}$) to $\sim 20$\,Myr at $Z\simeq 0.2\,Z_{\odot}$. 
The resulting average value of \xiion\ is then 
$\log\,$\xiion$/$\ergsHz\ $\simeq 25.68 \pm 0.13$.
This error includes the uncertainties of the EW(\CIII)s as well as those 
of the inferred ISM properties (\S \ref{ssec:UV_diagnostics}).
Likewise, the weaker LAEs sample is diagnosed to have 
a current star-formation age of $\sim 10$ to $500$\,Myr and 
$\log\,$\xiion$/$\ergsHz\ $\simeq 25.54 \pm 0.09$
if the binary stellar population models are assumed. 
The difference between the two samples arises primarily from the different \CIII\ EWs.
\begin{figure*}
	\centerline{
    		\includegraphics[width=0.85\textwidth]{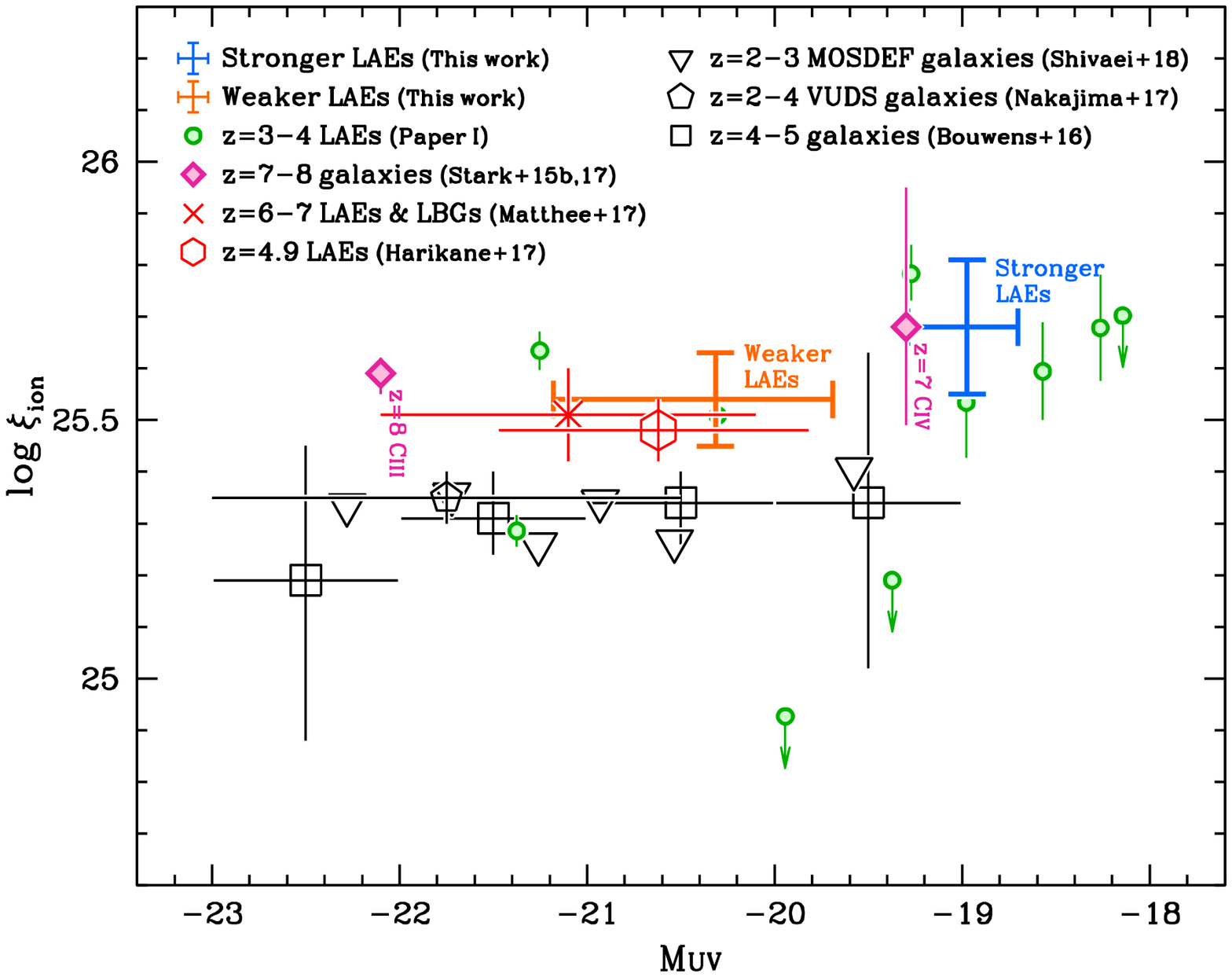}
  	}
  	\caption{
		The efficiency of ionising photon production \xiion\ 
		as a function of UV absolute magnitude. 
		Blue and orange refer to the stronger
		and the weaker LAEs samples respectively.		
		Green circles are individual LAEs whose \xiion\ were 
		derived from \Hb\ measures and recombination theory (Paper I). 
		Among the 13 LAEs presented in Paper I, one AGN-LAE and 
		two LAEs whose \xiion\ is less reliable are not displayed. 
		Magenta diamonds present a $z=7.05$ galaxy identified with \CIV\
		\citep{stark2015_c4}
		and a $z=7.73$ galaxy identified with \CIII\ \citep{stark2017}.	
		Red crosses represents a compilation of $z=6-7$
		luminous \Lya-selected and UV-selected galaxies 
		\citep{matthee2017}, and the red hexagon is an average for
		LAEs at $z=4.9$ \citep{harikane2017}.
		Black open symbols	show larger samples of continuum-selected 
		Lyman break galaxies at similar redshifts; 
		from MOSDEF (inverse triangles; \citealt{shivaei2018}; at $z=1.4-2.6$), 
		VUDS (pentagon; \citealt{nakajima2017}; at $z=2-4$; see also \citealt{lefevre2015,lefevre2017}), 
		and \citet{bouwens2016} (squares; at $z=3.8-5.0$).
		An SMC attenuation law is adopted in the correction for dust reddening. 
		\xiion\ values are calculated 
		under the assumption of a zero escape fraction of ionising photons excepting for those
		from MOSDEF which refer to an assumed value of 9\% (see text for details). 
	} 
	\label{fig:xi_MUV}
\end{figure*}

One uncertainty in interpreting \xiion\ for the weaker LAEs sample is whether 
or not to include binary stars in the stellar population. If we adopt the single star models, 
the star-formation age is significantly reduced to $\sim 1$ -- $7$\,Myr with 
$\log\,$\xiion$/$\ergsHz\ $\simeq 25.5$ -- $25.7$. The younger age arises from the fact 
that (i) single star models predict a smaller maximum EW of \CIII, and (ii) 
EW(\CIII) weakens more rapidly whereas binary evolution prolongs 
the period during which blue stars dominate the spectrum (see also \citealt{nakajima2017}). 
It is unlikely, however, that the weaker LAEs population is so young
(cf. \citealt{gawiser2006,gawiser2007,ono2010,guaita2011,kusakabe2018})
and indeed even younger than the stronger LAEs population. We note
that typical ages inferred for the LBG population at similar redshifts are 
several $\times 100$\,Myr (e.g. \citealt{reddy2012,nakajima2017}) where
binary star models are preferred. Since the weaker LAEs can logically be considered to 
represent a population intermediate between LBGs and the stronger LAEs
(e.g., \citealt{steidel2016,trainor2016,nakajima2017}), it seems reasonable to include binary stars.

The ISM properties and \xiion\ derived from the comparison of the UV 
emission lines with the photoionisation models of \citet{nakajima2017}
are given in Table \ref{tbl:properties_UV}. The uncertainties on \xiion
reflect both the variance arising from the use of stacked spectra and
the range of ages considered.

\begin{figure*}
	\centerline{
    		\includegraphics[width=0.9\textwidth]{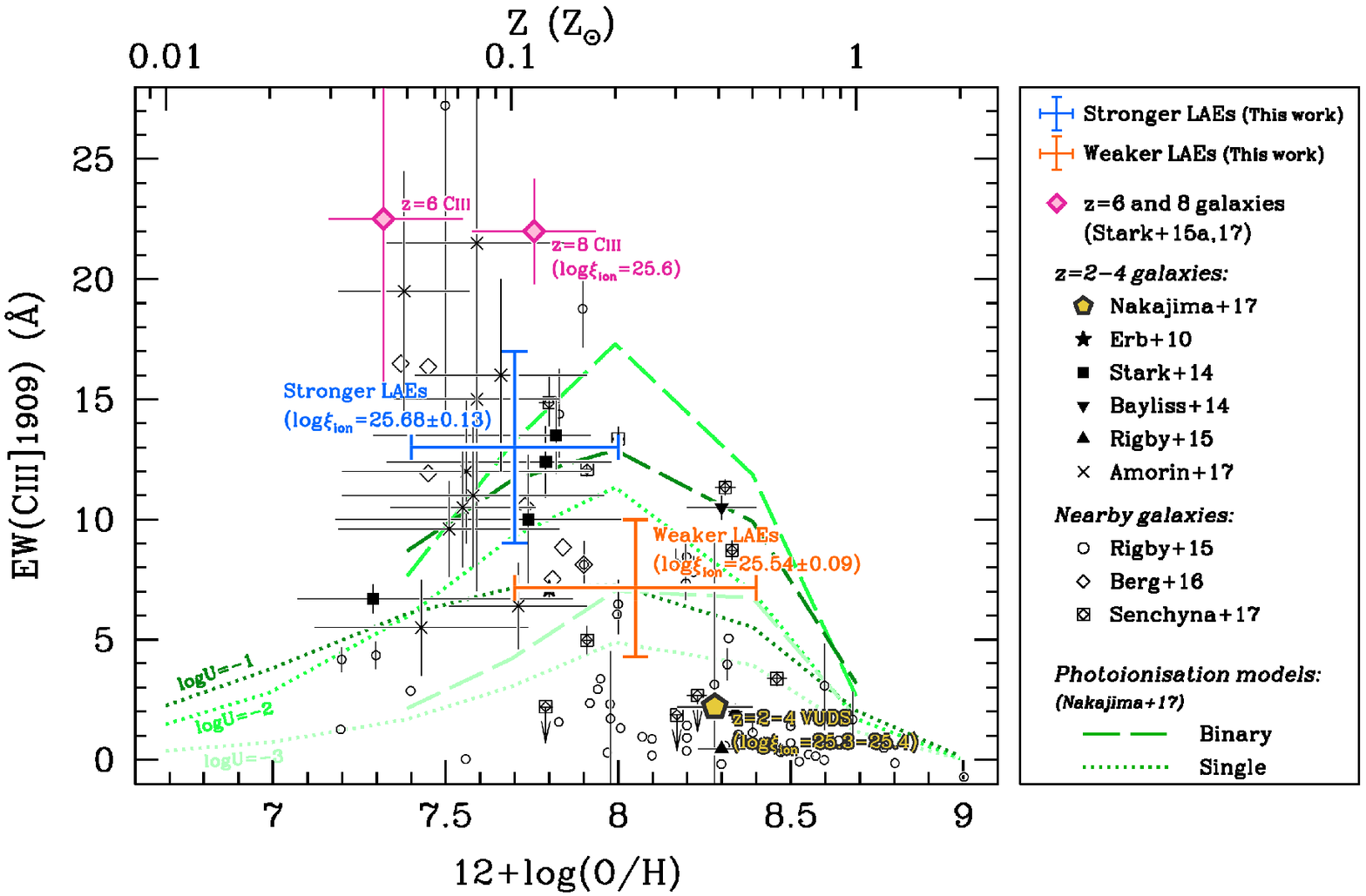}
  	}
  	\caption{
		The EW(\CIII) as a function of metallicity.
		The stronger and weaker LAEs sample 
		are indicated by the blue and orange error bars, respectively.
		The magenta symbol present the 
		$z=7.73$ galaxy identified with \CIII\ \citep{stark2017}, 
		and the black symbols represent $z=0-4$ 
		galaxies compiled from the literature as shown in the legend.
		The average LBG population at $z=2-4$ drawn from VUDS 
		\citep{nakajima2017} for which a \xiion\ measurement is available
		is highlighted in yellow.
		Green curves present photoionisation models 
		using single- (dotted) and binary- (long dashed) stellar 
		populations \citep{nakajima2017} and trace
		the maximum EW for a given metallicity and a range of ionisation 
		parameters: $\log U=-1$ (dark green), $-2$ (green), and $-3$ (light green).
	} 
	\label{fig:ewc3_Z}
\end{figure*}

\section{Discussion} 
\label{sec:discussion}

We present VIMOS measurements of the rest-frame UV nebular emission lines of \CIII, \CIV, \HeII, 
and \OIIIuv\ drawn from a sample of $70$ $z=3.1$ LAEs. By dividing the sample into nearly equal 
subsets based on the EW(\Lya), UV magnitude \MUV, and UV slope $\beta$, we demonstrate that 
the ``stronger LAEs sample'' with larger EW(\Lya), fainter \MUV\ and bluer $\beta$ have stronger 
EWs of \CIII\ emission and smaller (\CIII$+$\CIV)$/$\HeII\ line ratios than the corresponding 
``weaker LAEs sample''.
 
Using our photoionisation models and the UV diagnostic diagrams presented in \citet{nakajima2017}, 
we interpret these trends as indicating that the stronger LAEs sample, which represents the more 
intense LAE phenomenon, has typically a lower gas phase metallicity and, importantly, a more efficient
production of ionising photons. In addition to inferring a higher \xiion, binary stars are essential in our 
modelling to reproduce the EWs of \CIII.

We can place our results in context by examining, in Figure \ref{fig:xi_MUV}, the distribution of \xiion\
as a function of \MUV\ for the two subsamples of the VIMOS LAEs as well as for other high redshift 
sources, including the modest sample of LAEs studied with MOSFIRE in Paper I for which \xiion\ was 
inferred from \Hb\ using recombination theory. We also include various $z\simeq 5$ -- $8$ galaxies 
\citep{stark2015_c4,stark2017,matthee2017,harikane2017} for comparison purposes including several 
exploiting \CIII\ emission, and continuum-selected LBGs at $z=2-5$ 
\citep{bouwens2016,nakajima2017,shivaei2018}. 
The \xiion\ parameters of \citet{stark2015_c4,stark2017} and \citet{nakajima2017} are based on 
the UV emission lines with 
photoionisation models as we adopt in this paper, while the other studies estimate \xiion\ 
from hydrogen recombination lines.

Over the UV magnitude range from \MUV\ $\simeq -19.5$ to $-22.5$,
the continuum-selected LBGs at $z=2-5$ show 
a uniform value, $\log\,$\xiion$/$\ergsHz\ $\simeq 25.2$ -- $25.4$.
On the other hand, LAEs have a consistently larger
value by $\sim 0.2-0.3$\,dex, as originally claimed from a
much smaller sample in Paper I. Our analysis has also 
shown that UV faint LAEs present a higher \xiion\ than
their luminous counterparts.

The most intriguing comparison is with the emerging
data in the reionisation era. Although the uncertainties
remain large, and the samples modest in size, the large
values of \xiion\ inferred for the faintest metal-poor VIMOS LAEs
are similar to those observed in galaxies at $z>6$
\citep{stark2015_c4,stark2017,matthee2017,harikane2017}.
Since our LAEs likewise present intense \OIII$\lambda\lambda 5007,4959$ emission 
(Paper I) as inferred indirectly from \textit{Spitzer}/IRAC photometry for $z>6$ star-forming
galaxies \citep{smit2014,smit2015,oesch2015,roberts-borsani2016,harikane2017},
we conclude they remain valuable low-redshift analogues of the possible
sources of cosmic reionisation.

In order to understand the physical origin of the intense \CIII\ and
\Lya\ emission which combine to define the `stronger LAE sample'
as illustrated in Figure \ref{fig:ewc3_ewlya}, we present in Figure \ref{fig:ewc3_Z} the distribution of EW(\CIII) as a function of 
metallicity. Here we compare data from our two VIMOS-LAEs sub-samples 
with other relevant galaxies compiled from the literature. 

It is clear from Figure \ref{fig:ewc3_Z} that galaxies with a lower 
metallicity present a 
larger EW \CIII. One reason is that \CIII\ EW peaks at sub-solar metallicities, as 
shown with the photoionisation models. Another arises because the ionisation 
parameter becomes higher at lower metallicity. EW(\CIII) increases
with ionisation parameter for a given metallicity up to $\log U\simeq -2$.
As emphasised earlier, binary stellar populations can also produce 
stronger EWs of \CIII\ than single stellar populations at a fixed 
ISM properties \citep{nakajima2017}, although it seems likely
binary stellar populations are an ubiquitous feature of early
star-forming galaxies.

The other important parameter governing the EW(\CIII) 
is \xiion, or the age of the current star-formation. 
The photoionisation models in Figure \ref{fig:ewc3_Z} 
are based on $1$\,Myr old stellar populations providing a maximal 
EW(\CIII) for a given ISM condition. Weaker EWs 
correspond to longer star-formation ages and lower \xiion.

Finally we issue some caveats. The most important variable that could 
change our \xiion\ estimates is the escape fraction 
of ionising photons. EWs of \CIII\ and \CIV\ would be weakened with a higher 
escape fraction for a fixed ISM condition, since less ionising photons are necessary
to produce the high ionisation lines (see also \citealt{JR2016}).
For example, based on the models of \citet{JR2016}, for galaxies with a sub-solar 
metallicity, an ionisation parameter of $\log U\simeq-2.5$, and a $20$\,\%\ escape 
fraction would lower the EW of \CIII\ by a factor of 
$\lesssim 1.3$ as compared with the zero escape fraction case. However,
in the specific case of the MOSDEF survey of LBGs \citep{shivaei2018}
where an escape fraction of 9\% was typically assumed, applying a correction to our
adopted value of zero would only raise those points by $\simeq$0.04 dex in Figure \ref{fig:xi_MUV}
and not change our conclusions. Inevitably until we can independently examine the Lyman continuum
leakage from our sample, we cannot quantitatively break this degeneracy and
determine an absolute \xiion. 

A further issue is the fact our photoionisation models 
predict that the \CIII\ EW decreases if the 
metallicity becomes very much lower due to the lack of carbon.
Presumably, the positive correlation between EWs of \Lya\ and \CIII\ as shown 
in Figure \ref{fig:ewc3_ewlya} indicates such systems have not yet
been detected (see also \citealt{nakajima2017,harikane2017}). 

Furthermore the C$/$O abundance ratio could 
affect the interpretation. By default our photoionisation models
assume the empirical relationship between C$/$O and 
O$/$H ratios \citep{dopita2006} to predict the carbon lines strength. 
As discussed earlier, we have confirmed with (\CIII$+$\CIV)$/$\OIIIuv\
ratios that our VIMOS-LAEs typically have C$/$O ratios of 
$\log$ C$/$O $\simeq -0.7\pm 0.1$ and follow the empirical relation.
However, this may not be true for higher-$z$ galaxies. Indeed, the \CIII-or-\CIV\ 
emitting galaxies at $z=7-8$ studied by 
\citet{stark2015_c3,stark2015_c4,stark2017} have best-fit 
C$/$O ratios higher than expected from the empirical relation 
by $\sim 0.1$ -- $0.4$\,dex. Although the constraints are very weak for these
high-$z$ objects with the limited data, their large EWs of 
\CIII\ for their low metallicities, as presented in Figure \ref{fig:ewc3_Z},
would result from their elevated C$/$O ratios. The physical cause of such 
high C$/$O abundance ratios is unclear 
(see e.g., \citealt{mattsson2010,berg2016,nakajima2017} for a discussion).

In summary, therefore, the more intense LAEs with strong \CIII\
studied here and, by reference to Paper I, those with intense \OIII, can be considered to 
be young metal-poor galaxies in an early phase of galaxy evolution, 
providing a large amount of ionising photons into the ISM to achieve 
highly ionised nebular regions. They remain excellent analogues
of galaxies in the reionisation era and, by virtue of their
relative proximity and brightness, valuable targets for further
detailed study.

\bigskip
\noindent
{\bf Acknowledgements}

\noindent
We acknowledge financial support from European Research Council Advanced Grant FP7/669253 (TF, RSE). KN acknowledges a JSPS Overseas Research Fellowship. BER is supported in part by the program HST-GO-14747 provided by NASA through a grant from the Space Telescope Science Institute, which is operated by the Association of Universities for Research in Astronomy, Incorporated, under NASA contract NAS5-26555. It is a pleasure to thank Dan Stark, Masami Ouchi, Daniel Schaerer and the anonymous referee for useful comments and discussions. We also thank T. Hayashino, T. Yamada, Y. Matsuda and A.~K. Inoue for providing the LAE catalogue and the photometric data. The work is based on data products from observations made with ESO Telescopes at the La Silla Paranal Observatory under ESO programme ID 098.A-0010(A). Further data was taken with the W.M. Keck Observatory on Maunakea, Hawaii which is operated as a scientific partnership among the California Institute of Technology, the University of California and the National Aeronautics and Space Administration. This Observatory was made possible by the generous financial support of the W. M. Keck Foundation. The authors wish to recognise and acknowledge the very significant cultural role and reverence that the summit of Maunakea has always had within the indigenous Hawaiian community.  We are most fortunate to have the opportunity to conduct observations from this mountain.

\begin{table*}
  \centering
  \caption{Properties of the VIMOS-LAEs with UV-line detection
    }
  \label{tbl:properties_individuals}
  \renewcommand{\arraystretch}{1.25}
  \begin{tabular}{lcccccccl}
    \hline
    Obj. &
    R.A. &
    Decl. & 
    \MUV\ &
    EW(\Lya) &
    $z_{{\rm Ly}\alpha}$ &
    $\Delta v_{{\rm Ly}\alpha}$ &
    $\beta$ &
    UV line(s) \\
     &
    \tiny (1) &
    \tiny (1) &
    \tiny (2) &
    (\AA) \tiny (3) &
    \tiny (4) &
    (km\,s$^{-1}$) \tiny (5) &
    \tiny (6) &
    \,\,\tiny (7) \\
    \hline
    \multicolumn{9}{l}{\textit{-- NB497-selected --}} \\
    %
    LAE86177 & 22:16:47.6 & $+$00:17:51
     & $-19.3\pm 0.2$ & $111^{+33}_{-26}$ & $3.0734$ & -- & $-2.32^{+0.68}_{-0.82}$ 
     & \CIII, \OIIIuv$^{(\dag)}$ \\
    LAE104812 & 22:16:48.9 & $+$00:21:41
     & $-20.3\pm 0.1$ & $23^{+6}_{-5}$ & $3.0967$ & -- & -- 
     & \CIII, \OIIIuv$^{(\dag)}$, \HeII$^{(\dag)}$ \\
    LAE82902 & 22:16:59.6 & $+$00:17:16 
     & $-20.1\pm 0.1$ & $120^{+15}_{-13}$ & $3.0717$ & -- & $-1.50^{+0.34}_{-0.34}$ 
     & \CIII, \OIIIuv$^{(\dag)}$, \HeII$^{(\dag)}$ \\
    LAE104037 & 22:17:06.7 & $+$00:21:33 
     & $-21.4\pm 0.1$ & $36^{+2}_{-2}$ & $3.0673$ & $+194$ & $-1.62^{+0.08}_{-0.18}$ 
     & \CIII, \OIIIuv, \HeII\ \\
    LAE94460 & 22:17:08.0 & $+$00:19:32
     & $-19.9\pm 0.1$ & $51^{+8}_{-7}$ & $3.0744$ & $+171$ & $-1.96^{+0.40}_{-0.36}$ 
     & \CIII, \OIIIuv$^{(\dag)}$ \\
    LAE96549 & 22:17:26.3 & $+$00:19:58 
     & $-19.5\pm 0.2$ & $77^{+3}_{-3}$ & $3.0455$ & -- & $-2.44^{+1.26}_{-1.50}$ 
     & \CIII\ \\
    LAE84811 & 22:17:27.8 & $+$00:17:37
     & $-19.0\pm 0.3$ & $270^{+101}_{-69}$ & $3.0928$ & -- & $-1.20^{+1.04}_{-0.98}$ 
     & \CIII\ \\
    LAE46348 & 22:17:30.4 & $+$00:09:06
     & $-22.4\pm 0.1$ & $20^{+7}_{-7}$ & $3.1500$ & -- & $-0.96^{+0.04}_{-0.04}$ 
     & \CIII, \CIV ($+$abs.) \\
    LAE105870 & 22:17:33.3 & $+$00:21:51 
     & $-19.4\pm 0.2$ & $138^{+49}_{-35}$ & $3.0933$ & -- & $-3.76^{+1.46}_{-0.24}$ 
     & \CIII\ \\
    LAE52973 & 22:17:38.9 & $+$00:11:02
     & $-21.0\pm 0.1$ & $81^{+7}_{-6}$ & $3.0654$ & -- & $-1.42^{+0.10}_{-0.26}$ 
     & \CIII, \HeII$^{(\dag)}$ \\
    LAE104357 & 22:17:42.6 & $+$00:20:55 
     & $-21.8\pm 0.1$ & $97^{+3}_{-3}$ & $3.1732$ & -- & $-1.72^{+0.12}_{-0.10}$ 
     & \CIII, \CIV ($+$abs.), \HeII\ \\
    LAE65130 & 22:17:48.7 & $+$00:13:33
     & $-20.2\pm 0.1$ & $412^{+64}_{-53}$ & $3.0565$ & -- & $-1.34^{+0.36}_{-0.34}$ 
     & \CIII\ \\
    LAE52126 & 22:17:56.4 & $+$00:10:44
     & $-19.7\pm 0.2$ & $56^{+13}_{-11}$ & $3.1230$ & -- & $-1.18^{+0.44}_{-0.54}$ 
     & \CIII$^{(\dag)}$, \OIIIuv$^{(\dag)}$ \\
    AGN86861 & 22:17:09.6 & $+$00:18:01
     & $-21.5\pm 0.1$ & $82^{+3}_{-2}$ & $3.1097$ & $+338$ & $-2.38^{+0.08}_{-0.22}$ 
     & \CIV, \HeII, \NV, \HeII, \CIII\ \\
    \multicolumn{9}{l}{\textit{-- UV-selected --}} \\
    %
    LAE50416 & 22:17:08.1 & $+$00:09:58
     & $-21.1\pm 0.1$ & $117^{+3}_{-3}$ & $3.2911$ & -- & $-1.96^{+0.24}_{-0.28}$ 
     & \CIII, \OIIIuv, \CIV, \HeII$^{(\dag)}$ \\
    \hline
  \end{tabular}
  \\ 
  \vspace{-1mm}
  \begin{flushleft}
  	\small
	(1) Coordinates are in J2000.
	(2) Absolute UV magnitude.
	(3) Rest EW(\Lya). For the NB497-selected LAEs, the EW is estimated 
	from the BV$-$NB497 colour in conjunction with the \Lya\ redshift. 
	The EW of UV-selected galaxy of LAE50416 is derived from spectroscopy.
	(4) Redshift determined with \Lya. 
	(5) Velocity offset of \Lya. The systemic redshift is determined by
	the rest-frame optical emission lines provided by 
	the MOSFIRE observation (Paper I).
	(6) UV continuum slope $\beta$.
	(7) UV emission line(s) except for \Lya\ identified with VIMOS.
	($\dag$) Tentative detection. 
  \end{flushleft}
\end{table*}



\label{lastpage}

\end{document}